\documentclass[journal=jacsat,manuscript=article]{achemso}

\usepackage[version=3]{mhchem} 
\usepackage{physics}
\usepackage{subcaption}
\usepackage{hyperref}
\usepackage{color}
\usepackage{overpic}
\usepackage{bm}
\usepackage[utf8]{inputenc}

\newcommand{\bH}{\mbox{\protect\boldmath $H$}}

\newcommand{\bR}{\mbox{\protect\boldmath $R$}}
\newcommand{\bU}{\mbox{\protect\boldmath $U$}}

\def\AmS{{\the\textfont2 A}\kern-.1667em\lower.5ex\hbox
     {\the\textfont2 M}\kern-.125em{\the\textfont2 S}}

\def\AW{Addison\kern.1em-\penalty0\hskip0pt Wesley}
\def\BibTeX{{\rm B\kern-.05em{\smc i\kern-.025emb}\kern-.08em\TeX}}
\author{Jakub Martinka}
\affiliation{Heyrovsk\'{y} Institute, Academy of Sciences of the Czech \mbox{Republic, v.v.i.}, Dolej\v{s}kova 3, 18223 Prague 8, Czech Republic}
\alsoaffiliation{Department of Physical and Macromolecular Chemistry, Faculty of Science, Charles University, Hlavova 8, 128 43 Prague 2, Czech Republic}

\author{Mahesh Kumar Sit}
\affiliation{Heyrovsk\'{y} Institute, Academy of Sciences of the Czech \mbox{Republic, v.v.i.}, Dolej\v{s}kova 3, 18223 Prague 8, Czech Republic}

\author{Pavlo O. Dral}
\email{dral@xmu.edu.cn}
\affiliation[Second University]{State Key Laboratory of Physical Chemistry of Solid Surfaces, Department of Chemistry, College of Chemistry and Chemical Engineering, and Fujian Provincial Key Laboratory of Theoretical and Computational Chemistry, Xiamen University, Xiamen 361005, China}
\alsoaffiliation[Third University]{Institute of Physics, Faculty of Physics, Astronomy, and Informatics, Nicolaus Copernicus University in Toru\'n, ul. Grudziadzka 5, 87-100 Toru\'n, Poland}
\alsoaffiliation[Fourth Institution]
{Institute of Advanced Studies, Nicolaus Copernicus University in Toruń, ul. Wileńska 4, 87-100 Toruń, Poland}
\alsoaffiliation[Aitomistic]{Aitomistic, Shenzhen 518000, China}

\author{Ji\v{r}\'{i} Pittner}
\email{jiri.pittner@jh-inst.cas.cz}
\affiliation{Heyrovsk\'{y} Institute, Academy of Sciences of the Czech \mbox{Republic, v.v.i.}, Dolej\v{s}kova 3, 18223 Prague 8, Czech Republic}

\title[An \textsf{achemso} demo]
  {Toward Machine Learning Enhancement of Accelerated Surface Hopping with Scaled Spin--Orbit Couplings}

\abbreviations{UV}
\keywords{American Chemical Society, \LaTeX}

\begin{document}

\begin{tocentry}





	\includegraphics[height=4.45cm]{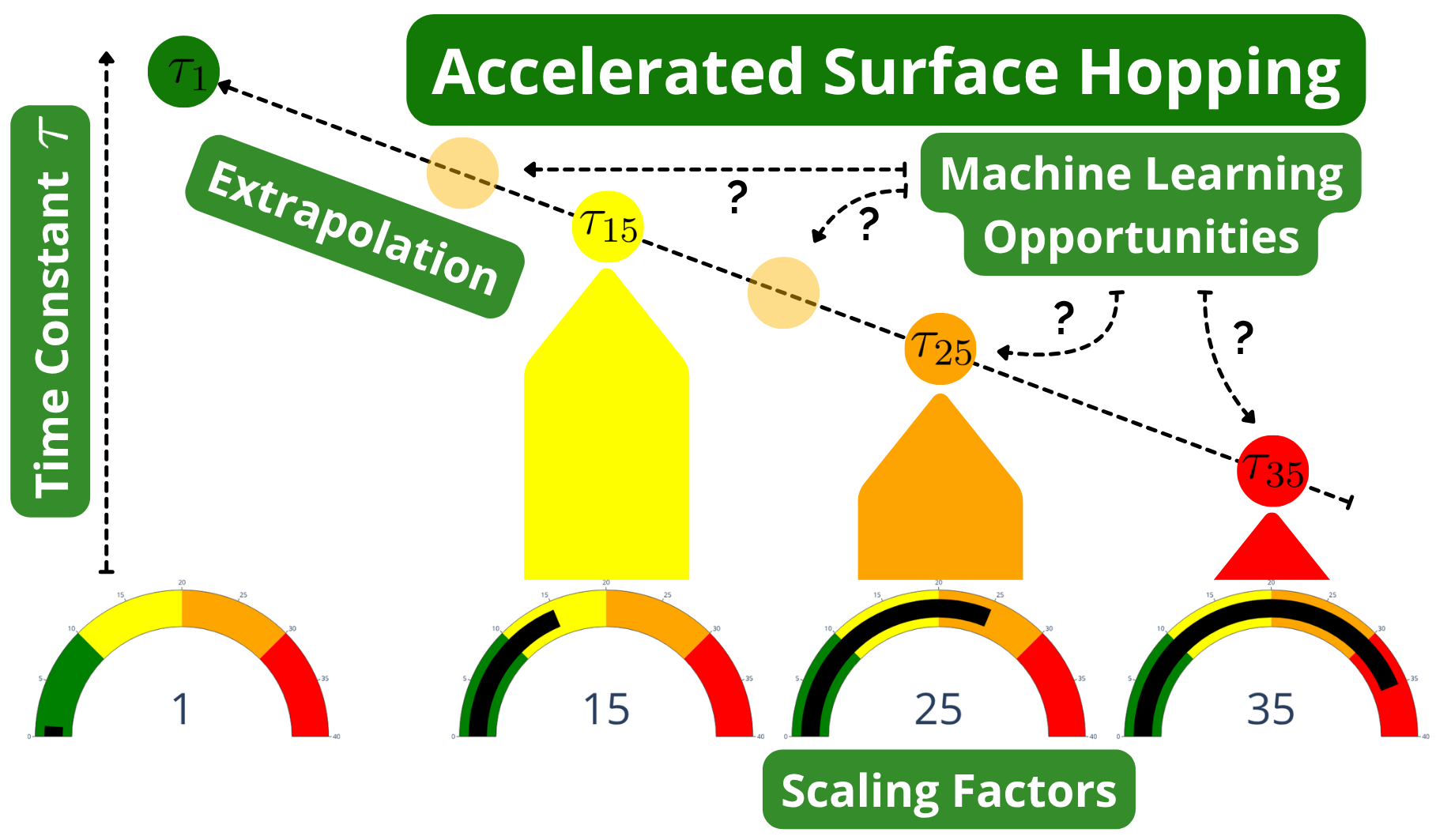}

\end{tocentry}

\begin{abstract}
Surface hopping (SH) methods are typically employed to simulate ultrafast nonadiabatic processes, but long timescales often remain beyond their reach.
To address this, accelerated SH scheme mitigates this limitation by scaling the driving forces of such process, either nonadiabatic couplings (NACs) in case of internal conversion or spin-orbit couplings (SOCs) for intersystem crossing.
However, obtaining the actual time constant requires extrapolation from several ensembles of trajectories with different scaling factors.
This introduces a significant computational demand, often restricting the number of trajectories per ensemble and, therefore, reducing the statistical confidence in the resulting time constant.
In this work, we investigate the accelerated scheme using silaethylene (CH$_2$SiH$_2$) as a case study, evaluating various population fitting methods and extrapolation techniques.
We trained machine learning models for potential energy surfaces (PESs) and NACs, and extended our rotate-predict-rotate approach to fit SOCs.
These models demonstrate high performance, yielding populations within the confidence interval of the reference MR-CISD/SA-CASSCF(2,2) data; however, the extrapolation itself is highly sensitive to the fitted time constants, leading to discrepancies in the final time constant.
Finally, we showcase and discuss how ML models can enhance the reliability of an accelerated SH scheme.
\end{abstract}


\section{Introduction}
Photochemical and photophysical phenomena play a fundamental role in governing how molecules interact upon the absorption of light, enabling the conversion of absorbed energy into chemical and physical transformations.
These processes are central to both natural systems and a wide range of technological applications, including photodynamic therapy\cite{Dolmans2003,Agostinis2011}, molecular solar thermal energy storage\cite{Wang2022}, and organic photovoltaics\cite{Bakulin2012,Lowrie2023}.
Most tools probing these phenomena within the nonadiabatic molecular dynamics (NAMD) aim to describe short timescales only.
This restriction is imposed by the need for high-accuracy excited-state potential energy surfaces (PESs), a demand that becomes even more significant when investigating larger chemical systems\cite{Dreuw2026}.
While hardware improvements and algorithm optimizations continue to push these limits, long timescales pose additional challenges, e.g. stability of integration methods or accuracy of NAMD algorithms, and available methods remain in an early phase of development\cite{Mukherjee2024a}.
However, if the objective of an NAMD study is to estimate a time constant or timescale, the couplings driving the process can be scaled to achieve acceleration\cite{Akimov2017,Pederzoli2019}; we refer to this as an accelerated scheme.

Within the variety of available NAMD methods, one of the most common is the trajectory surface hopping (SH), a member of the mixed quantum-classical family of algorithms that treats quantum and classical degrees of freedom separately\cite{CrespoOtero2018}.
The most common variant of SH was introduced in the 1990's by Tully\cite{Tully1990,HammesSchiffer1994} employing the fewest switches criterion, therefore abbreviated as FSSH.
Originally derived to describe internal conversion (IC) processes only, the approach was later extended to account for transition between states of different spin multiplicity mediated by spin-orbit couplings (SOCs), i.e., intersystem crossing (ISC) processes\cite{Takayanagi2002,Hu2007,Li2009,Abrahamsson2009,Fu2011,Han2011,Zaari2015,Curchod2016,Fedorov2016}.
Nevertheless, the extension of FSSH to ISC presents several problems.
For example, sum of hopping probabilities of all multiplets is not rotationally invariant, and when spin-orbit coupling is non-negligible, the PESs split, which must be accounted during dynamics propagation.
Granucci et al. demonstrated that each multiplet must be treated individually\cite{Granucci2012}; unfortunately, even simpler schemes based on Landau--Zener type of SH does not perform well in these cases\cite{Suchan2020}.
One solution to these concerns involves propagating the dynamics in a spin-adiabatic (i.e., diagonal) representation obtained by diagonalizing of the total Hamiltonian matrix (see Methods section).
However, this introduces numerical issues caused by arbitrary phases and unitary rotations among degenerate eigenvalues.
To address this, it has been proposed to use the unitary matrix, which diagonalizes the total Hamiltonian matrix, to transform the state coefficients from diagonal basis to spin-diabatic (i.e., molecular Coulomb Hamiltonian (MCH)) representation, and several techniques have been introduced to achieve this effectively\cite{Richter2011,Mai2015a,Pederzoli2017}.

\begin{figure*}[t!]
    \subfloat[]{%
        \begin{overpic}[width=.26\textwidth]{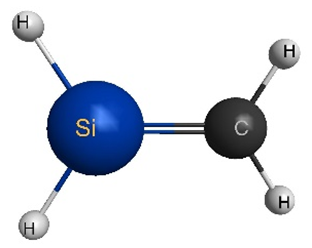}
        \put(18,7){\scriptsize 1}
        \put(20,68){\scriptsize 2}
        \put(80,12){\scriptsize 3}
        \put(80,60){\scriptsize 4}
    \end{overpic}
        \label{subfig:sich4}
    }\hfill
    \subfloat[]{%
        \includegraphics[width=.26\linewidth]{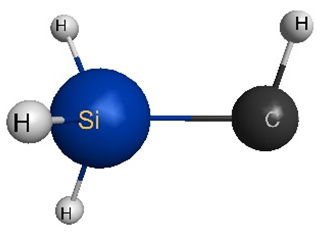}%
        \label{subfig:sih3ch}%
    }\hfill
    \subfloat[]{%
        \includegraphics[width=.26\linewidth]{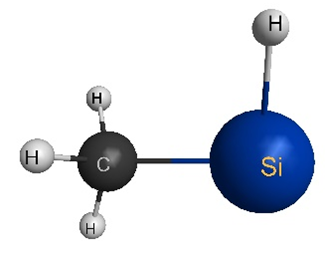}%
        \label{subfig:ch3sih}%
    }
    \caption{Optimized geometries of (a) CH$_2$SiH$_2$, (b) SiH$_3$CH and (c) CH$_3$SiH.}
    \label{fig:opt_fig}
\end{figure*}

In this work, we focus on silaethylene (CH$_2$SiH$_2$, see Fig.~\ref{subfig:sich4}), which has been studied previously both statically and dynamically, but never with inclusion of ISC relaxation to the triplet state.
At the optimized geometry, the $\pi\pi^*$ is the lowest singlet state; the Rydberg state $\pi$-3s is located only 0.6 eV above, but destabilized, and other states are excluded as they lie significantly higher in energy\cite{Pitonak2005}.
Thus, only the first two singlet states are included in the SH dynamics, following Ref.~\citenum{Zechmann2006}.
Silaethylene was studied at MR-CISD level-of-theory including two lowest singlet states within the FSSH scheme, where two major relaxation pathways, Type T and Type B, were identified.

We have recently compared the MR-CISD energies of silaethylene against Mukherjee's multireference coupled cluster theory incorporating triples perturbatively (MR-CCSD(T)), as well as the multireference averaged quadratic coupled-cluster (MR-AQCC) level-of-theory\cite{Plasser2025}.
Geometry optimization revealed a triplet ground state of SiH$_3$CH diradical (Fig.~\ref{subfig:sih3ch}) and a singlet ground state of CH$_3$SiH diradical (Fig.~\ref{subfig:ch3sih}), the latter of which is energetically close to optimized $S_0$ state.
Furthermore, structures related to rotation of the CH$_2$--SiH$_2$ bond are reported to be energetically degenerate.
Silaethylene is a small molecule, which allows for the use of a high level-of-theory for its description, while the small magnitude of SOCs (see Fig.~S9) indicates a slow transfer to the triplet state.
The estimation of time constant is, therefore, beyond the scope of standard FSSH including ISC, making it a perfect candidate for testing an accelerated FSSH scheme; specifically, by scaling the SOCs that drive the process.

In recent years, artificial intelligence (AI) has found a breathtaking applications in chemistry, ranging from protein structure predictions\cite{Jumper2021}, self-driving labs\cite{Leong2025}, chemical large language models\cite{Ramos2025}, up to AI agents autonomously performing computational chemistry simulations\cite{Zou2025,Hu2025}.
Apart from these exciting new developments, machine learning (ML) methods have been part of computational chemistry for many years, finding applications particularly in ground-state molecular dynamics (MD) simulations\cite{Behler2007,Behler2011,PhysNet2019,Schuett2018a,Bartok2010,Batatia2022}.
Their advantage lies in the fact that, once trained on a certain level of electronic structure data, the same accuracy is in principle achieved during propagation, but with a significant acceleration.
Furthermore, physical constrains can be encoded into the design of ML model, increasing both performance and robustness. 

Despite the great success of ML models for ground-state MD, describing excited-state events is far from simple, as excited PESs pose challenges even for standard electronic structure methods.
Electronic states become degenerate at conical intersections (CIs), where PESs become non-differentiable; a significant problem for ML methods that are designed to fit smooth functions.
Nevertheless, the use of ML methods within NAMD has become a research focus for several groups\cite{Dral2018,Dral2021,Westermayr2019,Akimov2021,Chen2018,Chen2019,Li2024a,Sit2023,Sit2024,Mausenberger2024a}.

In this work, we combine our previous efforts on learning vectorial properties and phase-correction workflows of learning NACs, applying them to fit SOCs using silaethylene as a test case.
This work demonstrates the generality of the Rotate-Predict-Rotate (RPR) scheme as well as the learning procedure of targets with arbitrary phases.
This approach significantly accelerates NAMD simulations involving slow ISC to the triplet manifold.
We conduct an analysis of accelerated FSSH scheme and critically discuss the opportunities of ML application within an accelerated SH schemes.

\section{Methods}
\subsection{Trajectory surface hopping involving spin-orbit effects}
To account for ISC, one must include spin-orbit effects in the Hamiltonian.
The total Hamiltonian $\hat{H}^{\text{tot}}$ is expressed as the sum of the molecular Coulomb Hamiltonian (MCH) 
\begin{equation}
    \hat{H}^{\text{MCH}}=-\sum_i\frac{1}{2m}\nabla^2_i+\frac{e^2}{4\pi\epsilon_0}\left( \sum_{A<B}\frac{Z_AZ_B}{R_{AB}} -\sum_i\sum_A\frac{Z_A}{R_{iA}}+\sum_{i<j}\frac{1}{r_{ij}} \right),
\end{equation}
and the Breit-Pauli spin-orbit Hamiltonian,
\begin{equation}
    \hat{H}^{\text{SO}}=\frac{1}{2m^2c^2}\left( \sum_i\sum_A \frac{Z_Ae^2}{r^3_{iA}}\hat{\bm l}_{iA}\cdot\hat{\bm s}_i - \sum_{ij}\frac{e^2}{r_{ij}^3}\hat{\bm l}_{ij}\cdot(\hat{\bm s}_i+2\hat{\bm s}_j) \right).
\end{equation}
The matrix elements of this Hamiltonian between components of spin multiplet states can be expressed using the Wigner-Eckart theorem in terms of reduced matrix elements which form an axial Cartesian SOC vector\cite{SOC1977}.
We apply ML to the SOC vector, using the RPR procedure to maintain its covariance (improper rotations are not included in RPR, so the axial character does not matter).
In particular, the singlet-triplet SOC matrix elements are identical to the components of the SOC vector, but this approach can be generalized to treat arbitrary spin multiplicity.

In the Tully's FSSH scheme, the coefficients in MCH basis $\{\phi_i\}$ are propagated according to
\begin{equation}\label{eq:coeff}
    \frac{\text{d}c_k(t)}{\text{d}t}=\sum_j(-iH_{kj}^{\text{tot}}-\sigma_{kj})c_j(t),
\end{equation}
where $\sigma_{kj}={\dot{\bR}(t)}\cdot\bra{\phi_k}\nabla_{\bm R}\ket{\phi_j}$ is the time-derivative coupling, expressed as the scalar product of the velocity and the nonadiabatic coupling (NAC), and $H_{kj}^{\text{tot}}$ is the matrix element of the total Hamiltonian, which is typically diagonal.
However, Granucci et al. demonstrated that the transition probabilities in the MCH representation are not rotationally invariant\cite{Granucci2012}.
Furthermore, when SOCs are non-negligible, the dynamics should be performed in the basis of the eigenvectors $\{\psi_i\}$ of the total Hamiltonian, i.e., diagonal representation.
This involves using a unitary matrix $\bm U$ to diagonalize the total Hamiltonian matrix and to transform the quantities in Eq.~\ref{eq:coeff}.
Unfortunately, this is not trivial due to the occurance of $\text{d}{\bm U}/\text{d}t$ in the equations of motion, which causes numerical instabilities.
Solutions to this problem have been proposed, such as the 3-step propagator approach by Gonz\'{a}lez et al.\cite{Mai2015a}, or the removal of redundant degrees of freedom from $U$ before taking the derivative by some of us\cite{Pederzoli2017}.
In this work, we employ the former approach, which transforms the coefficients in the diagonal basis $\bar{c}_i$ to the MCH basis as
\begin{equation}
    c_i = \sum_jU_{ij}\bar{c}_j,
\end{equation}
followed by propagation in the MCH basis analogously to Eq.~\ref{eq:coeff}, and finally back-transformation to the diagonal representation.
A hop from state $k$ to $j$ is performed according to probability $p$ given by
\begin{equation}
p(k\rightarrow j) = \left(1-\frac{|\bar{c}_k(t+\Delta t)|^2}{|\bar{c}_k(t)|^2}\right) \times \frac{\text{Re}(\bar{c}_j(t+\Delta t)P^*_{jk}\bar{c}^*_k(t))}{|\bar{c}_k(t)|^2-\text{Re}(\bar{c}_k(t+\Delta t)P^*_{kk}\bar{c}^*_k(t))},
\end{equation}
where $P_{ij}$ is an element of the propagator matrix defined as
\begin{equation}
    P_{ij}(t+\Delta t, t) = \bU^\dagger(t+\Delta t)\text{exp}(-\Delta t(i\bH^\text{tot}-{\bm \sigma}))\bU(t).
\end{equation}
A hop is initiated when a uniformly generated random number $r\in[0,1]$ satisfies the condition
\begin{equation}
    \sum_{i=1}^{j-1}p(k\rightarrow i)<r\leq\sum_{i=1}^jp(k\rightarrow i).
\end{equation}

\subsection{Accelerated surface hopping scheme}\label{subsec:ash}
The most computationally demanding step in SH is solving the electronic Sch\"{o}dinger equation, which inherently limits NAMD to simulating phenomena at hundreds of femtoseconds timescales, depending on the size of the system, basis set, and level of theory.
While long timescales have been probed for different NAMD schemes in recent years\cite{Akimov2017,Westermayr2019,Li2020,Akimov2021,Li2021a,Mukherjee2022a,Mukherjee2024a}, trajectory-based methods can suffer zero-point energy leakage when using Wigner-based initial conditions\cite{Mukherjee2022}.
Fortunately, if the goal of an NAMD study is to determine the time constant or lifetime, it is possible to use an accelerated SH scheme.
This is based on artificially scaling the couplings that drive the process of interest by a constant $\alpha>1$, i.e., ISC driven by SOCs as
\begin{equation}\label{eq:scaling}
    \hat{H}_{ij}^\alpha = \alpha\bra{\phi_i}\hat{H}_\text{SO}\ket{\phi_j}.
\end{equation}
The selected $\alpha$ values should be small enough not to violate the short-time approximation; however, very small $\alpha$ values may lead to difficulties in fitting the population dynamics.
The appropriate choice is system-specific, as it is influenced by the magnitude of the couplings and the energetic proximity of the states.
In practise, they are typically selected empirically by running several trajectories to determine whether ISC takes place.

This accelerated scheme is justified by the fact that, in the weak-coupling regime, the transfer rate is well described by Marcus theory\cite{Marcus1993,Marcus1956} (or equivalently by Fermi's golden rule or nonadiabatic transition state theory\cite{Lorquet1988}, among others), which predicts that, for a two-state model, the rate constant $k$ depends quadratically on the coupling.
By scaling the couplings, the time constant $\tau$ scales with $\alpha^{-2}$, 
\begin{equation}\label{eq:tau2}
    \tau^\alpha = \frac{A}{\alpha^2},
\end{equation}
which we refer to as the \texttt{Fit1} method throughout the text.
However, Eq.~\ref{eq:tau2} holds only for a two-state system; the general $N$-state relationship is not known; when scaling NAD coupling in Ref.~\cite{Nijamudheen2017} the authors assumed a rational fraction form with additional constant parameter in the denominator. We did not employ this fitting formula as it would lead to nonzero ISC rate for SOC scaled to zero.

The time constant or lifetime is obtained by extrapolation to $\alpha=1$, which requires running multiple ensembles of trajectories (possibly with different initial conditions) for a set of scaling constants to determine $A$ and $B$.
In a typical accelerated SH study, we are unfortunately limited to using only a few different scaling factors.
Eq.~\ref{eq:tau2} can be extended by introducing an additional parameter, $\tau_\infty$, so that the model reproduces the expected limiting behaviour, with the lifetime diverging as $\alpha\rightarrow 0$ and approaching the finite value $\tau_\infty$ as $\alpha\rightarrow\infty$,
\begin{equation}
    \tau^\alpha = \frac{A}{\alpha^2} + \tau_\infty.
\end{equation}

Another approach is to fit the data on a log--log scale using Eq.~\ref{eq:tau2}, i.e., using a linear function $C\alpha+D$, where $D$ corresponds to $\log A$ and the slope $C$ is optimized but expected to be near $-2$, consistent with Eq.~\ref{eq:tau2}.
This approach is referred to as the \texttt{Fit2} method and, unlike \texttt{Fit1}, does not restrict the exponent to be $-2$, although, as is the case of $\tau_\infty$, the additional fitting parameter generally increases the uncertainties and widens the confidence intervals.

\subsection{Machine learning driven trajectory surface hopping}
In supervised learning, an ML model maps input features, typically represented by a molecular descriptor, to targets, i.e., molecular property of interest.
There is a variety of molecular descriptors, commonly divided into global types, e.g. relative-to-equilibrium (RE)\cite{Dral2017}, Coulomb matrix (CM)\cite{Li2021,Westermayr2020a} and others, or local, e.g. smooth overlap of atomic positions (SOAP)\cite{Bartok2013} and others.
Furthermore, one can combine a descriptor derived from molecular geometry with an additional property if it is available and correlates with the target.

The choice of ML regressor is another non-trivial task, as models are generally categorized into groups of neural networks and kernel methods, each containing different architectures.
We limit ourselves to well-established approaches within the MLatom package\cite{MLatom2024}, specifically the multi-state ANI\cite{Martyka2025} (MS-ANI) model and kernel ridge regression (KRR) with Gaussian kernel.
We point the reader to the relevant reviews discussing the strengths and weaknesses of various ML approaches (e.g. Refs.~\citenum{Dral2020,Noe2020,Westermayr2020c,Pinheiro2021,Mueller2025}).

Data set construction is a crucial part of any successful ML application.
An active learning (AL) procedure,  based on uncertainty quantification, iterative data augmentation and model retraining, might be an ideal solution, but currently we do not have deep integration with the NAMD code with ISC adaptation, which is one of our future goals. In addition, AL still requires manual experimentation such as careful selection of system-specific thresholds, although progress has been made in this direction\cite{Martyka2025,Westermayr2019}.
Once reference trajectories are available, a training set may be created by random selection, assuming that all the relevant PES regions and geometries are well represented.
Rather than building the training set from full trajectories, selecting snapshots from across the entire data pool may be preferred\cite{Tiefenbacher2025}.

ML-driven FSSH accounting for IC and ISC relaxation requires the PESs and their gradients, as well as NACs and SOCs.
While fitting PESs is a well-established task for which robust AL schemes exists\cite{Martyka2025,Westermayr2019,Bispo2025}, NACs and SOCs possess different challenges.
NACs are vectorial, diverge at CI, and can have an arbitrary sign due to the wave-function phase.
Conversely, SOCs are typically smooth, delocalized and can be complex-valued; but are also vectorial and can have arbitrary sign.
Learning vectorial targets is problematic because ML models must satisfy rotational equivariance.
Several approaches have been proposed to address this, such as explicitly building equivariance into the model architecture\cite{Schutt2021,Grisafi2018,Wilkins2019} or using RPR approach developed by some of the authors of this study\cite{Martinka2024}.
The latter has proved to be a simple and efficient technique to fit dipole moments and polarizabilities, and NACs in the context of NAMD\cite{Martinka2025}.
It relies on training the model on a pre-aligned data set; during the inference stage, the molecular structure is rotated to a reference frame before the ML prediction, and the resulting vectorial property is then back-rotated to the original orientation.
Training is further complicated by arbitrary sign in the training database; these can be preprocessed before training\cite{Westermayr2019,Martinka2025}, or alternativelty, a phase-less loss can be employed in the case of neural networks\cite{Westermayr2020b}.
Additionaly, commonly used descriptors were designed for learning PESs and might perform poorly for other properties~\cite{Martinka2025}.

\section{Results}
\subsection{Reliability of the Accelerated Surface Hopping Scheme}
Before discussing opportunities for ML applications, we first examine the accelerated SH scheme based on quantum chemical method for silaethylene, involving two singlet states ($S_0$, $S_1$) and one triplet ($T_1$).
We use the same settings for static calculations as in Ref.~\citenum{Plasser2025}, specifically MR-CISD/SA-CASSCF(2,2) with aug-cc-pVDZ basis set in COLUMBUS (version 7.2, 2022)\cite{Columbus7.2,Lischka2020}, which we abbreviate to MR-CISD further in the text for simplicity.
The 500 initial conditions for the dynamics were generated using the Wigner distribution of a quantum harmonic oscillator\cite{Barbatti2010b} for each normal mode from the optimized ground-state geometry at 298 K.
From these, 200 trajectories were propagated for each of the scaling factors 15, 25 and 35, i.e., 600 trajectories in total.
Once the ML models were trained, the ensembles were extended to cover all 500 initial conditions to generate 1,500 trajectories in total.
In the case of ML trajectories, we also tested independent sets of initial conditions (500 trajectories per $\alpha$ value; see Fig.~S14) to mitigate potential bias, as well as different numbers of trajectories for each scaling factor ($\alpha=15: 700, \alpha=25: 600, \alpha=35: 500$; see Fig.~S15) to increase the confidence in the results obtained for smaller $\alpha$ values.
These tests may be relevant for assessing the robustness of the accelerated SH scheme; however, we do not find any significant discrepancies that would affect the conclusions reported for the same set of initial conditions.

We employed our ISC adaptation of Newton-X\cite{Pederzoli2017,NewtonX2014,NewtonX2026}, with all trajectories initiated in $S_1$.
The final time of the dynamics was set to 500 fs, with a nuclear time step of 0.5 fs and an integration step for the electronic Schr\"{o}dinger equation of 0.025 fs.
The dynamics were performed using 3-step propagator approach\cite{Mai2015a} in the diagonal representation.
The electronic state coefficients were transformed to the MCH basis to obtain corresponding populations.
Following a successful hop, the momenta were rescaled along the NACs to ensure conservation of the total energy.
In case of a frustrated hop, the momenta were left unchanged.
No decoherence correction was applied because existing decoherence schemes have primarily been developed for systems with rapidly decaying NACs, whereas their applicability to dynamics involving significant SOCs remains uncertain\cite{Granucci2012,Heller2021}.
In the plots showing populations, the average of $|c_i|^2$ is plotted; its validity was confirmed by comparing it against the singlet and triplet composition of the active PES, averaged over all trajectories, as well as against dynamics with phase correction (see Fig.~S25).
Out of 200 MR-CISD trajectories, an average of 12 terminated shortly before the end of the propagation due to inconsistent energies or transition densities reported by Columbus.
The ML-driven trajectories were free of such failures.

While accelerated NAMD schemes have been applied to both NACs\cite{Nijamudheen2017,Lingerfelt2016} and SOCs\cite{Pederzoli2019,WasifBaig2021,WasifBaig2025}.
The primary objective of this work is to study the ISC process; we therefore apply the scaling factor exclusively to SOCs.
The SOCs facilitate the transition to the triplet state, represented by the increase in its average adiabatic population (see Fig.~\ref{subfig:mrci_pop}).
This population is summed over all triplet components and fitted using $A^\text{pop}(1-\exp (-t/\tau^\alpha))$, where the $A^\text{pop}$ constant corresponds to the asymptotic maximum triplet population.
This choice is rationalized by the plateau in $T_1$ population for sufficiently large $\alpha$ (see Fig.~S8).

Rather than fitting each ensemble of trajectories independently, we employ a multi-curve fitting strategy that enforces the constant $A^\text{pop}$ to be identical across data corresponding to different scaling factors $\alpha$.
The multi-curve fitting improves the numerical stability of the fits and robustness of the overall extrapolation part of the accelerated scheme (see Fig.~S7 without multi-curve strategy).
For each population fit, we perform 5000-sample bootstrapping\cite{Efron1979,Mooney1993,Efron1994} by randomly selecting $n$ trajectories with replacement to evaluate the uncertainty associated with the fitted parameters.
The resulting values are then used to define confidence intervals for the extrapolated time constants.

\begin{figure*}[t!]
    \subfloat[\texttt{Fit1} method.]{%
        \includegraphics[width=.48\linewidth]{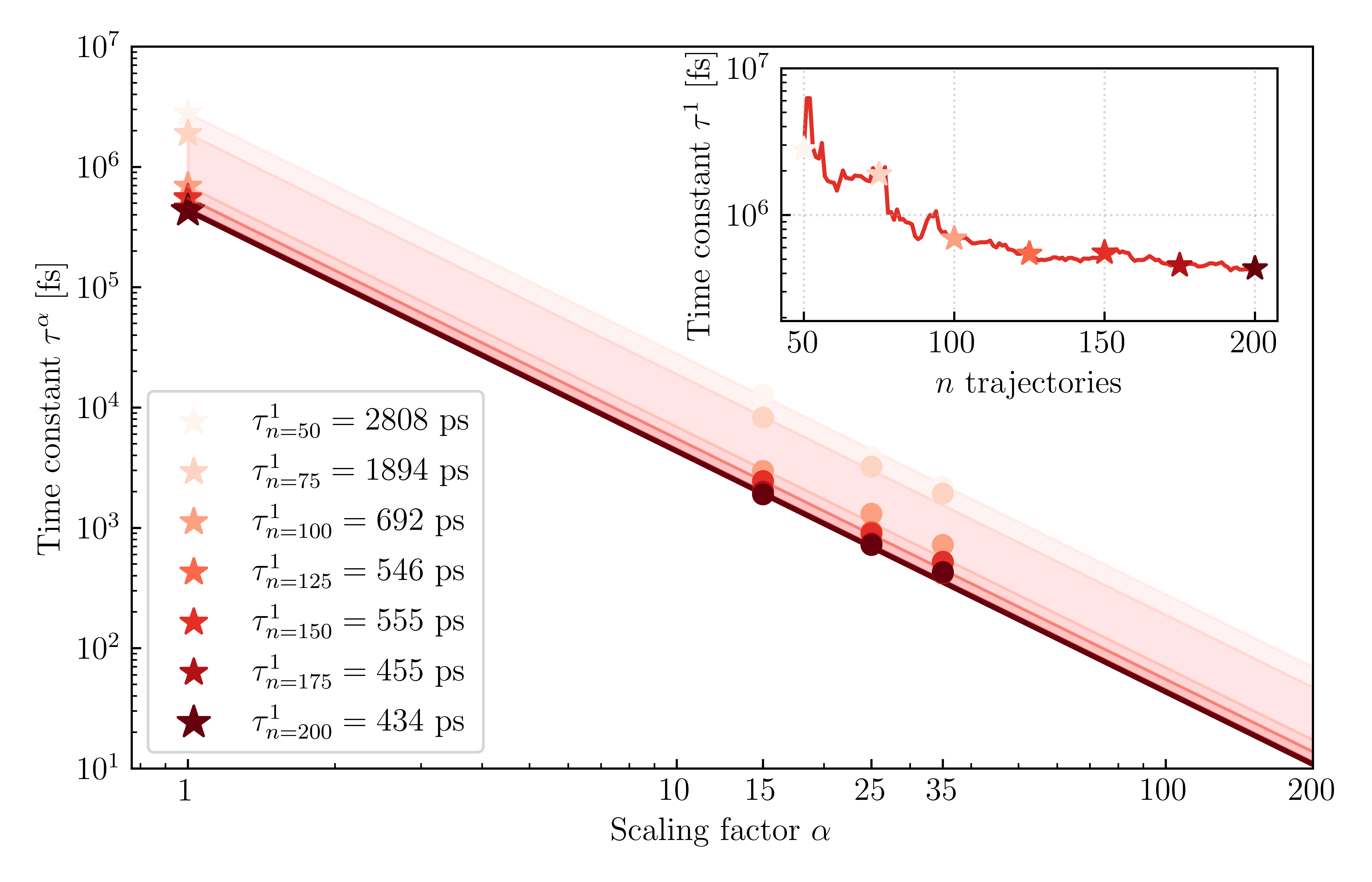}%
        \label{subfig:mrci_npop1}%
    }\hfill
    \subfloat[\texttt{Fit2} method.]{%
        \includegraphics[width=.48\linewidth]{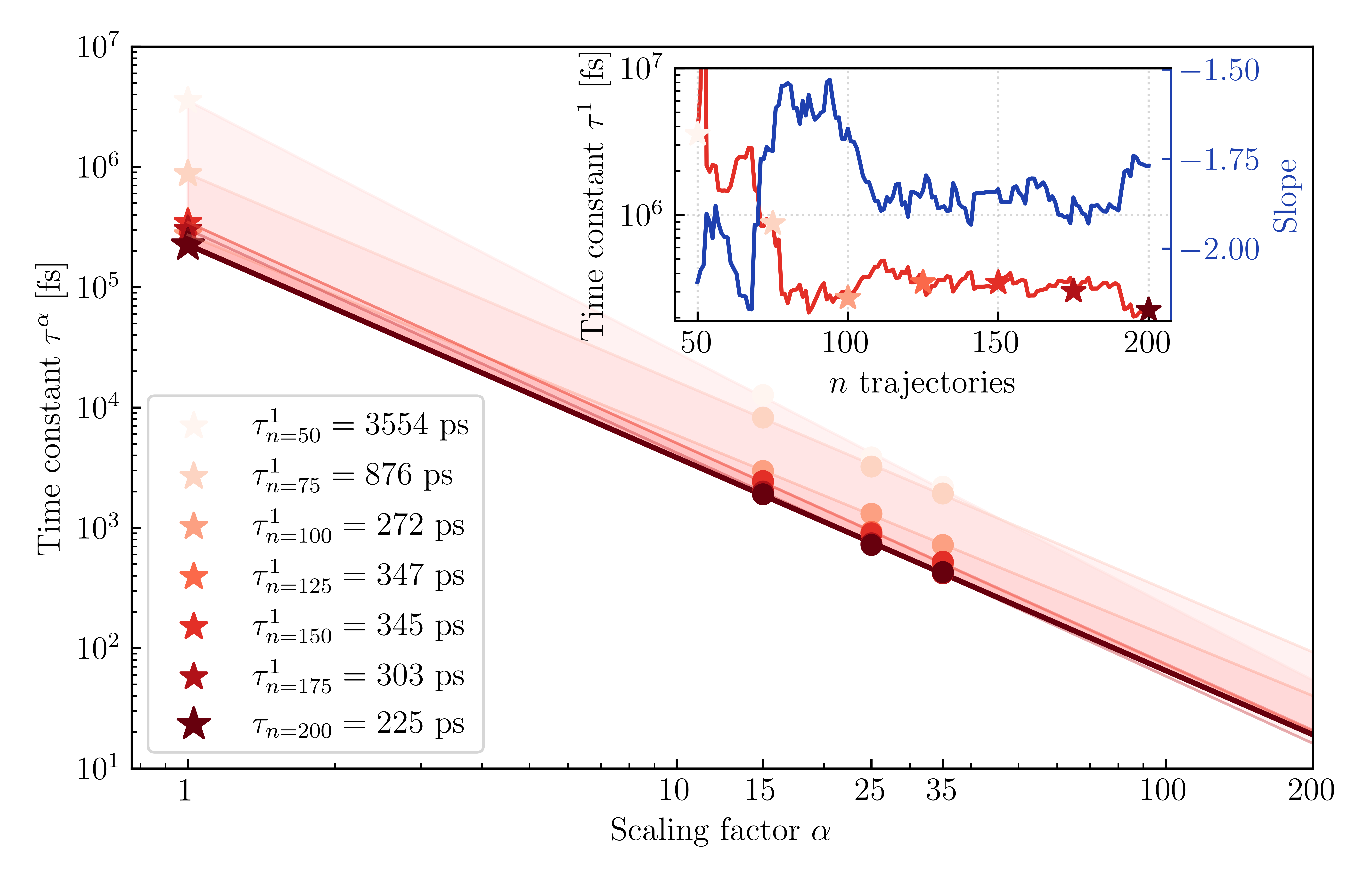}%
        \label{subfig:mrci_npop2}%
    }
    \caption{Extrapolation of MR-CISD time constants to $\alpha=1$ using \texttt{Fit1} and \texttt{Fit2} methods, based on the number of trajectories $n$. The top-right plot shows the time constants obtained by analyzing $n$ trajectories; in the case of \texttt{Fit2}, the slope is shown in blue. Scaling factors of 15, 25, and 35 were used.}
    \label{fig:mrci_npop}
\end{figure*}

The use of an accelerated SH scheme with scaling of couplings inherently requires several ensembles of trajectories to obtain confident population fits for each scaling factor.
These fits provide time constants that can be used for extrapolation to a scaling factor of 1 using the \texttt{Fit1} and \texttt{Fit2} methods.
We also compare the approach using $\tau_\infty$ in the case of 500 ML trajectories (see Fig.~S12), as well as for the 200 MR-CISD trajectories (see Fig.~S13).
For $N$ ensembles, $N$ times more trajectories must be propagated, which significantly increases the computational demand; consequently, the number of trajectories $n$ within each ensemble is typically sacrificed in standard accelerated SH studies.
Using three ensembles of MR-CISD trajectories, we analyzed the uncertainty introduced by the number of trajectories $n$ in each ensemble.
Fig.~\ref{fig:mrci_npop} illustrates the extrapolation to scaling factor of 1 based on different values of $n$, ranging from 50 to 200 trajectories.
The extrapolated time constants using the \texttt{Fit1} (Fig.~\ref{fig:mrci_npop}a) and \texttt{Fit2} (Fig.~\ref{fig:mrci_npop}b) methods are compared.
The top-right plot shows the extrapolated time constants obtained from an analysis of the first $n$ trajectories; specific values (in ps) are displayed in the bottom-left.
In the \texttt{Fit1} method, where the exponent is fixed to the theoretical value of $-$2, the extrapolated time constants decreases and slowly converges to 434 ps for an ensemble of 200 trajectories.
The \texttt{Fit2} method, which fits points in log--log scale and does not restrict exponent, converges to 226 ps for 200 trajectories.
The slope ranges from $-$1.5 to $-$2.1 (the blue curve on top right of Fig.~\ref{fig:mrci_npop}b), but once at least 110 trajectories are included in the analysis it fluctuates around slope value of $-$1.84.
Regardless of the fitting method, our analysis suggests that at least 100 trajectories per ensemble should be included to obtain reliable estimate of the time constant at $\alpha=1$.
Same conclusion is obtained while analyzing trajectories based on ML models (see Fig.~S3).

The requirement for multiple ensembles, each comprising hundreds of trajectories, presents an opportunity for ML to increase confidence of SH schemes accelerated by scaled couplings.
We first briefly discuss approaches that turned out problematic and were thus not followed and analyzed in further detail.
One natural idea for ML application is to increase $N$ by generating more ensembles with different scaling factors $\alpha$, thereby providing more points for extrapolation to 1.
However, this proved difficult for smaller values of $\alpha$ due to the numerical challenge of fitting shallow, slowly increasing triplet populations.
Furthermore, extending such trajectories beyond the reference timescale (500 fs in our case) may cause instabilities in the dynamics due to extrapolation outside the training set; this is strongly system- and process-specific and might eventually be resolved by advanced sampling technique.
On the other hand, using ensembles with larger $\alpha$ is not recommended, as it may negatively influence the multi-curve fit and the extrapolated time constants.
Finally, increasing $N$ within the existing range of scaling factors, e.g. using \{15, 20, 25, 35\} instead of \{15, 25, 35\}, does not significantly affect the resulting time constant when compared to the uncertainty related to a small $n$ (see Fig.~S1).
We therefore conclude that the most promising use of ML in an accelerated SH scheme is to expand the size of the trajectory ensembles themselves.

\subsection{Evaluating Accuracy of Machine Learning Models}
When applying ML models in production simulations, their performance is typically evaluated on an independent, hold-out test set, unseen during model training or selection.
To construct this data set, we used 100 MR-CISD trajectories with scaling factor 20, generating a randomly selected pool of 6,000 points, of which 5,000 were used for training and 1,000 for testing.
For silaethylene, considering two singlet states and one triplet state, the model must describe three adiabatic energies, corresponding gradients, one NAC between singlet states and two SOC elements.

The most critical aspect is the accurate learning of PESs, as they govern the dynamics and thus determine the exploration of the configurational space.
We trained an MS-ANI model for the $S_0$ and $S_1$ PESs, while the $T_1$ surface was fitted using a separate ANI model on the same training set.
All models achieved satisfactory performance on the test set, with $R^2$ values exceeding 0.95 for $S_1$, 0.98 for $S_0$, and 0.99 for $T_1$.
Therefore, a crucial part of the evaluation involves focusing on the NAMD simulations themselves and their trajectory analysis.

In case of silaethylene, the conical intersection is reached through the torsion around C--Si bond\cite{Pitonak2005}.
Furthermore, the C--Si stretch and symmetric bipyradimidalization were previously reported as the main modes playing role in the $S_1$ deactivation\cite{Zechmann2006}.
Once the symmetric bipyramidalization is the main motion driving the relaxation, the trajectory is called type B.
When the torsion around the C--Si dominates, the trajectory is regarded as type T.
These two modes can be characterized by additional coordinates, $\theta$ and $\beta$ defined as

\begin{subequations}
\begin{equation}
    \theta = \sqrt{\frac{\theta_1^2+\theta_2^2}{2}},
\end{equation}
\begin{equation}
    \beta = \sqrt{\frac{\beta_1^2+\beta_2^2}{2}},
\end{equation}
\end{subequations}
where $\theta_1$ (H$_1$-Si-C-H$_3$) and $\theta_2$ (H$_2$-Si-C-H$_4$) dihedral angles capture the central torsion, whereas, $\beta_1$ (C-H$_1$-H$_2$-Si) and $\beta_2$ (Si-H$_3$-H$_4$-C) describe the relative orientation of the methylene and silylene groups (see Fig.~\ref{subfig:sich4}).

\begin{figure}
     \centering
     \begin{subfigure}[b]{0.49\textwidth}
        \centering
        \includegraphics[width=\textwidth]{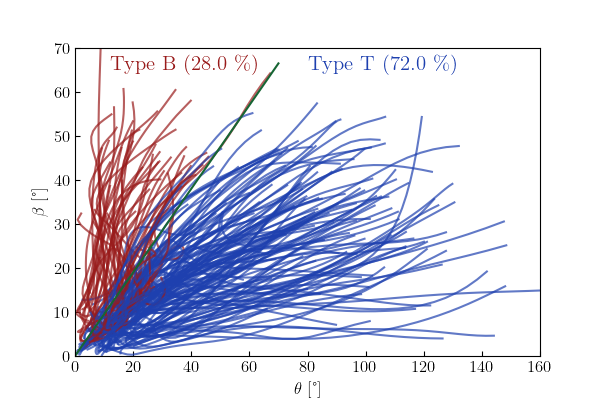}
    \end{subfigure}
    \hfill
    \begin{subfigure}[b]{0.49\textwidth}
        \centering
        \includegraphics[width=\textwidth]{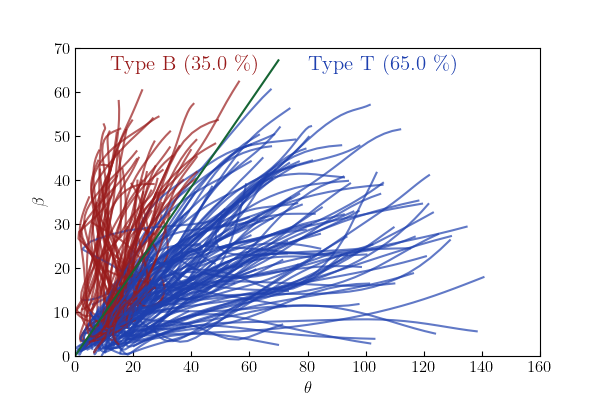}
    \end{subfigure}
    \caption{Projection of MR-CISD (left) and ML (right) trajectories with scaling factor $\alpha=15$ onto $\beta-\theta$ plane up to 20 fs. The deactivation process of silaethylene is well recovered by ML-driven trajectories. Scaling factor of SOCs does not influence this projection in both cases (see Fig.~S4 and Fig.~S5).}
    \label{fig:tb}
\end{figure}

The projection of MR-CISD and ML trajectories onto the $\beta-\theta$ plane provides an ideal test for the ML-PESs.
These two modes exert a strong influence on the dynamics during the early stage, and the trajectories can be shown to be separated by the $\beta=0.95\cdot\theta$ line (see Fig.~\ref{fig:tb}).
Type T trajectories are characterized by simultaneous torsional and Si--C stretching.
The latter lifts the degeneracy, causing the trajectories to hop later compared to the type B trajectories, which are driven by the Si--C stretch into the region of the bipyramidalized crossing seam\cite{Zechmann2006}.
Type B trajectories reach the crossing seam faster than type T, meaning that the ratio between the two modes must be reproduced by the ML models to achieve agreement with the reference $S_1$ population decay.
As evident from Fig.~\ref{fig:tb}, the MR-CISD and ML trajectories agree well when projected onto the $\beta-\theta$ plane up to 20 fs.
In case of the 200 MR-CISD trajectories, 72~\% trajectories are classified as type T and 28~\% as type B.
In comparison, the ML trajectories contain 65~\% type T and 35~\% of type B.
This difference is caused by trajectories that land in proximity to the $\beta=0.95\cdot\theta$ line within 20~fs, which alters the classification ratio.
However, both distributions are close to the previously reported ratio of 68~\% and 32~\%\cite{Zechmann2006}.
This indicates that the $S_1\rightarrow S_0$ deactivation channel is well recovered, which will be demonstrated on $S_1$ decay in Fig.~\ref{subfig:mrci_pop}.

The NACs between $S_0$ and $S_1$ were fitted using KRR with RPR approach and the gradient difference as a descriptor.
Prior to fitting, we followed the established protocol of multiplying the NACs by the energy gap $\Delta E_{01}$ to remove non-smooth behavior\cite{Westermayr2020a} and applied phase-correction to the data set using iterative procedure described in Ref.~\citenum{Martinka2025}.
SH simulations with the NAC model are expected to be sensitive to the minute changes in the model quality and, hence, we adopt a special model selection strategy based on the lessons for similar cases from the a previous work by one of us.~\cite{Ge2024} The key lesson is that among many  models with small errors in the hold-out test set, some perform very well in simulations and some very badly. Hence, test errors are insufficient and a proxy, cheap-to-compute property, which is connected to what is important in the final simulations, can be selected to evaluate the model quality before running the simulations. In our case, such proxy-property can be the error in NACs, 
in the nonadiabatic regions (where NACs are non-zero), i.e., for small $\Delta E_{01}$.
To address this while maintaining simplicity, we constructed a dedicated 'selection set' with 1,000 points from the 5,000-point pool by taking only the geometries with smallest $\Delta E_{01}$, yielding a diverse set of configurations with a relatively small energy gap, which is complementary to and not overlapping with the hold-out test set.
The training set was generated by repeatedly sampling 5,000 points using different random seeds, followed by training separate ML models on each subset.
The models were evaluated on the 1,000-point selection set, and the best-performing one was retained for subsequent dynamics simulations.

We verified that R$^2$ exceeding 0.9 was achieved on the hold-out test set. However, as mentioned above, such static metrics are not decisive and do not necessarily reflect the performance in the production simulations\cite{Ge2024}.
A more reliable assessment requires evaluating the model within dynamical simulations, where preliminary results indicate satisfactory performance.

The SOCs were trained following a procedure similar to that used for the NACs, utilizing RPR with phase-correction.
However, in contrast to NACs, SOCs are smooth and typically slowly varying, which eliminates the need for multiplication by the energy gap and requires the use of a different descriptor.
We benchmarked several rotationally invariant descriptors, such as RE, CM, SOAP; however, the best performance was unexpectedly obtained using Cartesian coordinates translated to the center of mass and rotated into a standard orientation via RPR based on tensor of inertia (for detailes see Ref.~\citenum{Martinka2024}).
This somewhat counterintuitive result is consistent with our previous experience, where an RE descriptor augmented with selected Cartesian coordinates performed best for dipole moment prediction.
As with the NAC model, such comparisons should be interpreted with caution, as the ultimate validation lies in reproducing reference dynamics.
While the model predicting SOCs between $S_0$ and $T_1$ achieved good performance (R$^2$ over 0.98) on the randomly selected test set, comparable accuracy could not be obtained for the $S_1-T_1$ SOCs.
Nevertheless, the $S_1-T_1$ SOCs are extremely small, and setting them to zero along the entire trajectory resulted in an indistinguishable population dynamics.
This indicates that their contribution is negligible and that the reduced accuracy of the corresponding model does not affect the overall dynamical outcomes.

\subsection{Machine Learning Models Reproducing Reference Trajectories}
\begin{figure*}[t!]
    \subfloat[Populations obtained from 200 MR-CISD and 200 ML trajectories with scaling factors of 15, 25, and 35.]{%
        \includegraphics[width=.95\textwidth]{figures/populations_sich4.png}
        \label{subfig:mrci_pop}
    }\\
    \subfloat[Extrapolation of MR-CISD time constants using \texttt{Fit1} and \texttt{Fit2} methods to scaling factor $\alpha=1$.]{%
        \includegraphics[width=.48\linewidth]{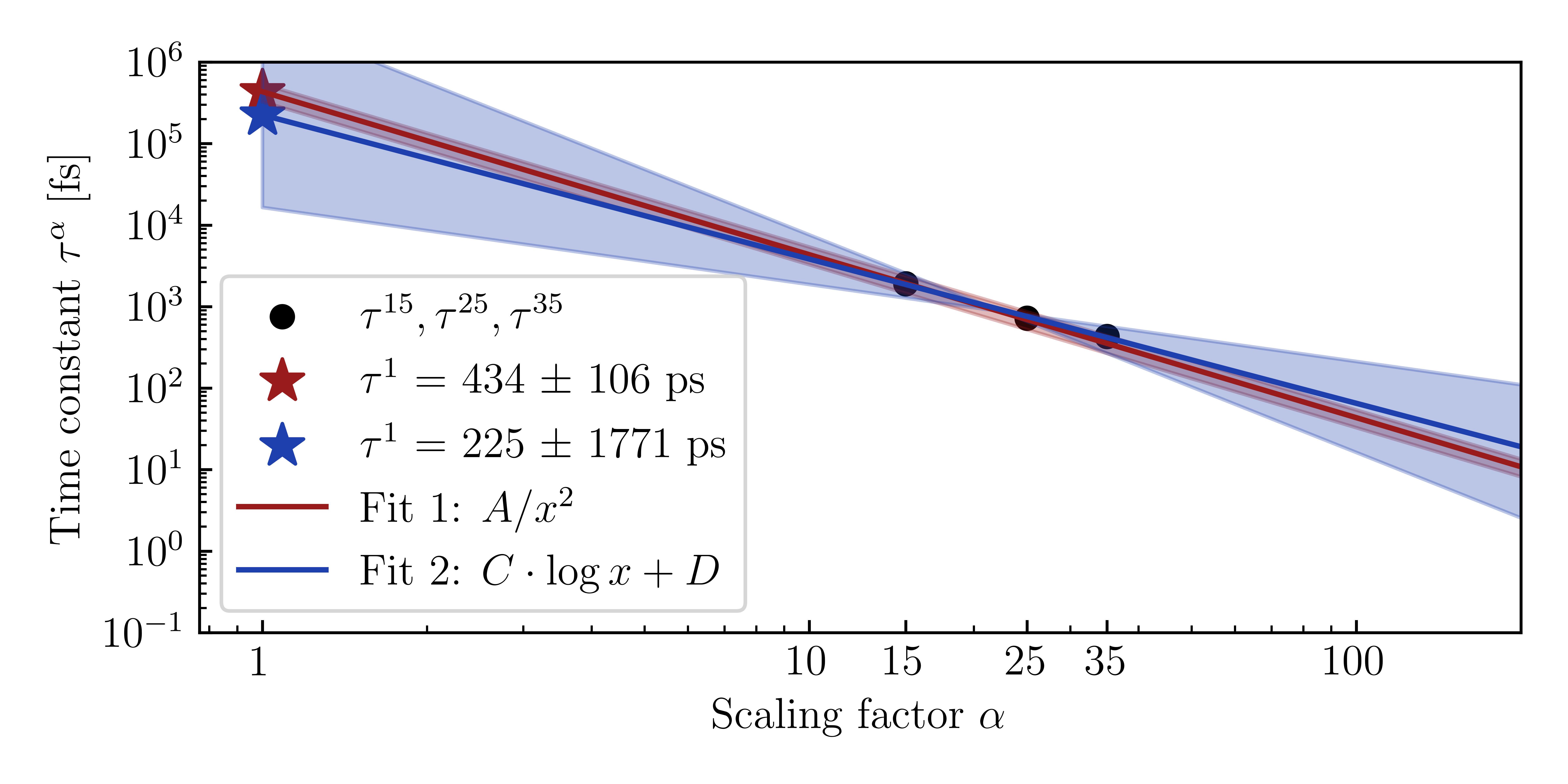}%
        \label{subfig:mrci_extr}%
    }\hfill
    \subfloat[Extrapolation of ML time constants using \texttt{Fit1} and \texttt{Fit2} methods to scaling factor $\alpha=1$.]{%
        \includegraphics[width=.48\linewidth]{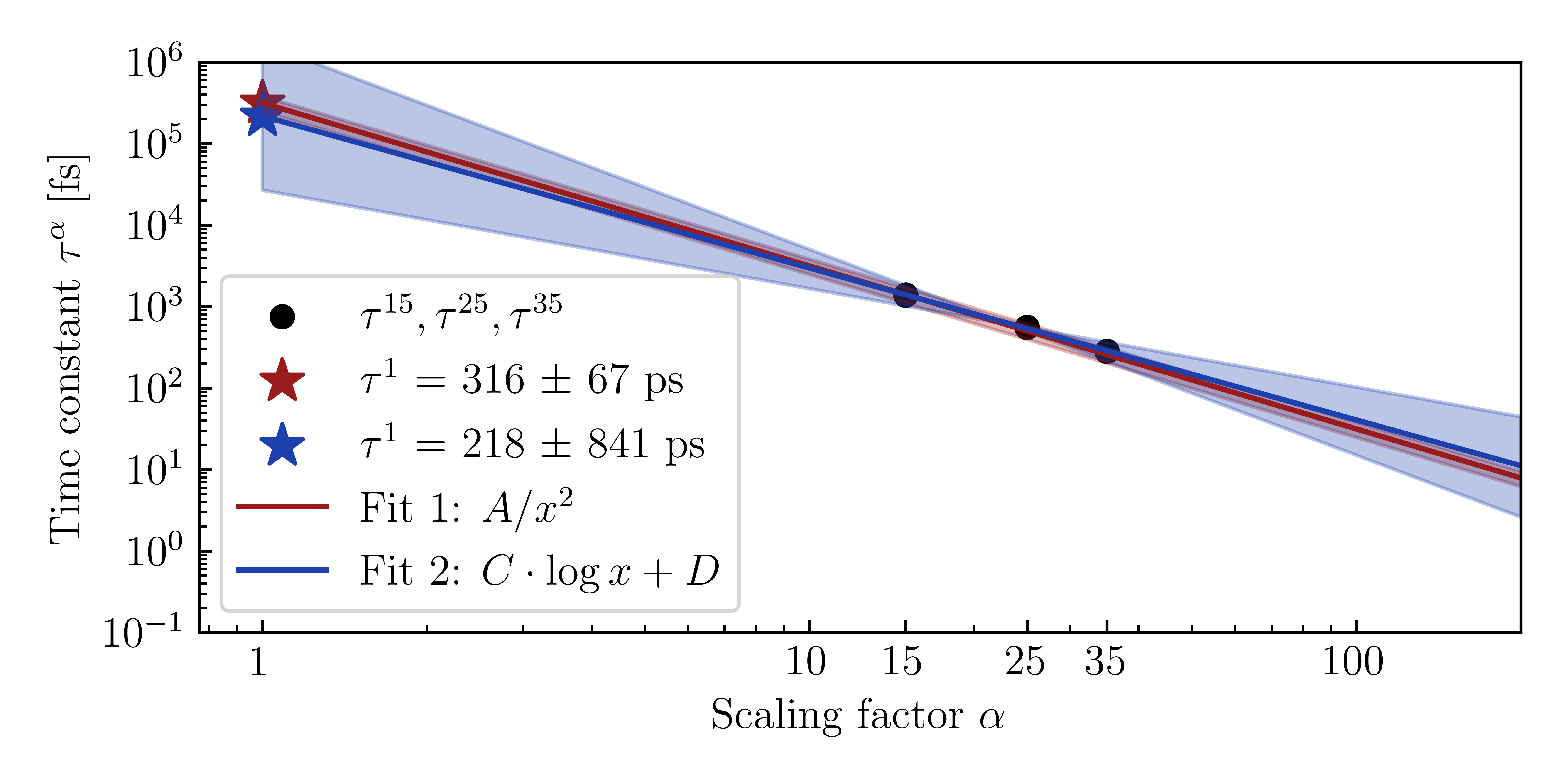}%
        \label{subfig:ml_extr}%
    }
    \caption{
    Comparison of MR-CISD and ML results from FSSH simulations of silaethylene. All trajectories were initiated in $S_1$, and the deactivation to $S_0$ is well reproduced. The transition to $T_1$ falls within the confidence interval (a). However, the $T_1$ fits are highly sensitive, leading to discrepancies when extrapolated to a scaling factor of 1 (b, c).}
    \label{fig:fig2}
\end{figure*}

The final test of the ML models is their performance within SH simulations facilitated by the ISC-adapted Newton-X interfaced with MLatom, using the same simulation parameters and initial conditions as the MR-CISD reference.
The resulting ML populations are plotted in Fig.~\ref{subfig:mrci_pop} (dark lines), alongside the MR-CISD populations (light lines) discussed previously.
By investigating the state coefficients in the MCH basis from the 200 MR-CISD trajectories for all $\alpha$ values (15, 25 and 35) we collected configurations related to changes of the $T_1$ state within an interval of 0.5 to 0.95.
It can be seen from individual trajectories that ISC is mainly described as $S_0\rightarrow T_1$, while majority of selected configurations are associated with a CH$_2$--SiH$_2$ angle of 90 degrees (see Figs.~S16-S18), small $S_0-T_1$ energy difference (see Figs.~S19-S21), and relatively large SOCs (see Figs.~S22-S24).
Furthermore, the slow ISC is explained by an energy barrier associated with accessing configurations with large SOC (see Figs.~S12 and S13).

The overall population transfer is well-described by the ML models, given the relatively simple procedure used to assemble the training sets.
The $S_1$ decay is in excellent agreement with the MR-CISD reference, indicating that the ML-PESs and ML-NACs achieved satisfactory performance.
However, it is apparent that the ML populations are not as smooth as the reference; this could be improved by increasing the number of trajectories $n$ within an ensemble; the primary motivation for using ML models, which are significantly faster than MR-CISD for trajectory propagation.
Furthermore, after 150 fs the $S_1$ state remains more populated in the ML results, suggesting potential for further refinement.
This could potentially be resolved through an AL; however, such an approach requires a complex integration of ISC-adapted NAMD and ML codes, which is beyond of scope of this work.

Despite the ML $T_1$ populations falling within the reference MR-CISD error margin $\epsilon (t)$ of the 95\% confidence interval calculated as
\begin{equation}\label{eq:confidence}
    \epsilon (t) = 1.96\sqrt{\frac{p_i(t)(1-p_i(t))}{N_\mathrm{traj}}},
\end{equation}
the multi-curve fits result in MR-CISD time constants of a $1903\pm 581$, $722\pm 260$ and $429\pm 175$ fs, compared to $1381\pm 363$, $555\pm 151$, $283\pm 82$ fs for the ML models at scaling factors 15, 25 and 35, respectively.
These time constants are used to extrapolate to $\alpha=1$ for both MR-CISD (see Fig.~\ref{subfig:mrci_extr}) and ML (see Fig.~\ref{subfig:ml_extr}).
For the \texttt{Fit1} method, the MR-CISD gives time constant of $434\pm 106$ ps, while ML gives $316\pm 67$ ps.
For the \texttt{Fit2} method, MR-CISD yields $225\pm 1771$ ps and ML gives $218\pm 841$ ps.
While the extrapolation using \texttt{Fit2} method agrees closely, this is likely a coincidence resulting from the different values of $\tau^\alpha$ and the fitted slopes ($-$1.86 for ML and $-$1.77 for MR-CISD, both of which are near the theoretical value of $-$2).
The \texttt{Fit1} method more closely highlights the difference in the data points, as the fixed exponent leads to a more pronounced divergence in the extrapolated values.

The discrepancy in the time constants is mainly caused by the ensemble of trajectories corresponding to scaling factor 35.
This is evident if we replace the ML population with the MR-CISD reference, effectively assuming the trajectories at $\alpha=35$ were perfectly reproduced by the ML models (see Fig.~S6a).
In such a scenario, the recomputed ML time constants would be $1762\pm 369$, $742\pm 154$, and $407\pm 85$ fs, leading to extrapolated time constants of $407\pm 68$ ps and $190\pm 420$ ps for the \texttt{Fit1} and \texttt{Fit2} methods (see Fig.~S6b), respectively, which are closer to the reference values of $434\pm 106$ ps and $225\pm 1771$ ps.
One possible interpretation is that ML errors are amplified by the scaling factor, leading to noisier and more fluctuating population transfer for larger values of $\alpha$.
This observation also argues against using ML to increase the number of trajectory ensembles $N$ by adding points with large scaling factors $\alpha$.
Given the evidence, we recommend using a hybrid approach: running one set of trajectories with the largest scaling factor, then using the resulting data to train ML models that are subsequently employed with smaller scaling factors.
This approach would benefit from MR-CISD data within multi-curve fitting, while the ML models would save a substantial amount of time.
However, the current results suggest that ML models should be applied with special care within an accelerated SH scheme, given the high sensitivity of the extrapolated results to the fitted time constants.

\begin{figure}[H]
    \centering
    \includegraphics[width=.95\textwidth]{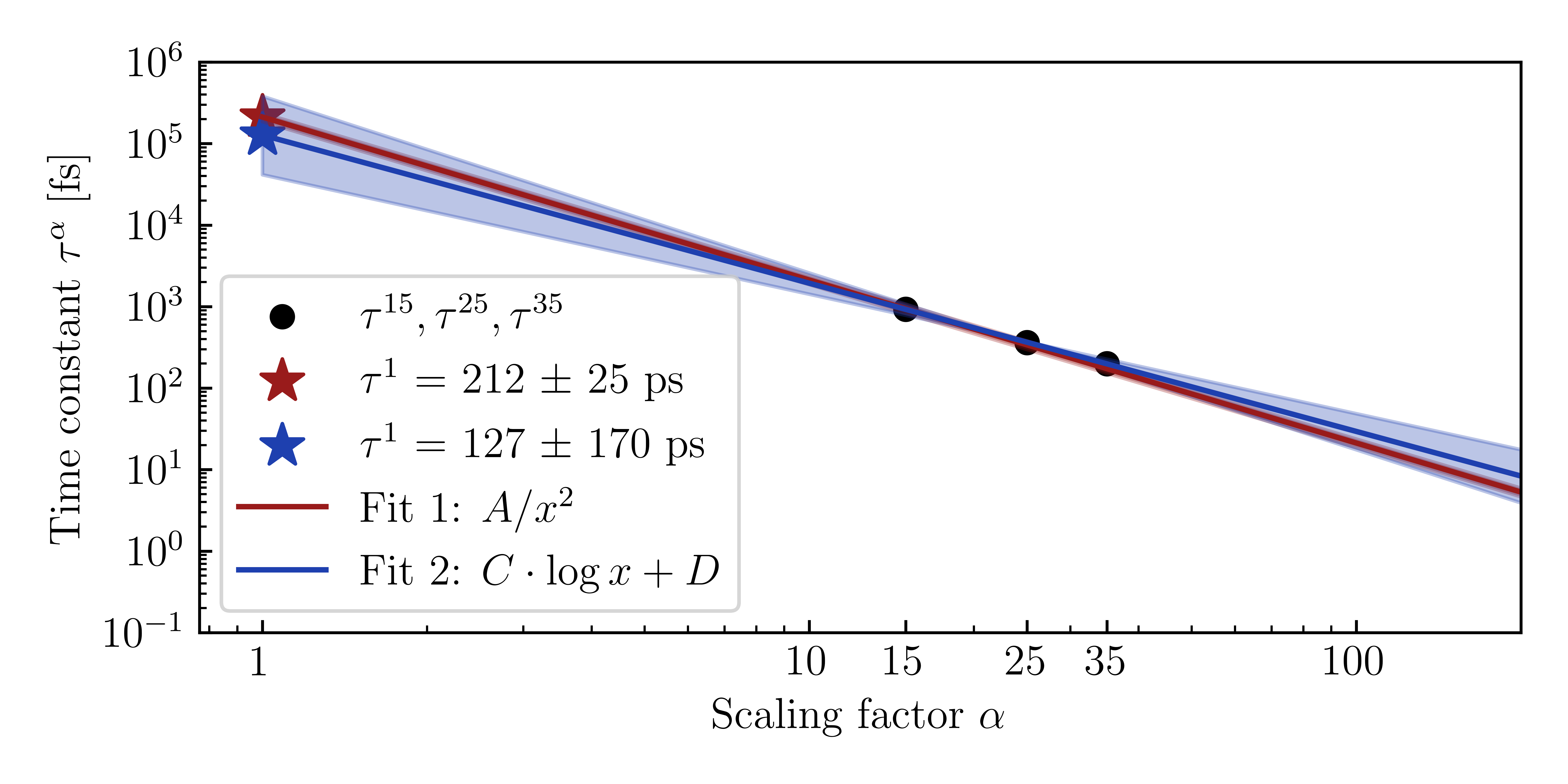}
    \caption{Extrapolation of ML time constants using \texttt{Fit1} and \texttt{Fit2} methods to scaling factor $\alpha=1$ obtained from 500 ML trajectories with scaling factors 15, 25, and 35. The confidence intervals for the extrapolated values are obtained by applying the same extrapolation formula using $\tau\pm$ the 5000-sample bootstrap error.}
    \label{fig:extrpo_500}
\end{figure}

The reported time constants have large bootstrapping errors, which are amplified during the extrapolation.
Fortunately, ML makes it possible to run much larger trajectory ensembles.
In our case, we performed an additional 300 trajectories, increasing the total number to 500 (see Fig.~\ref{fig:extrpo_500} and Fig.~S2; same applies to Figs.~S14--S15), although significantly larger ensembles should become feasible once the entire procedure is optimized.
For the larger ensemble, the bootstrapping errors are substantially reduced, yielding time constants of $928 \pm 173$, $363 \pm 53$, and $199 \pm 30$ fs, and extrapolated time constants of $212 \pm 25$ ps and $127 \pm 170$ ps for the \texttt{Fit1} and \texttt{Fit2} methods, respectively.
Although these results cannot be directly compared with the MR-CISD reference, they demonstrate that ML can substantially improve the reliability of the accelerated SH method by enabling much larger trajectory ensembles.

\section{Conclusions and Outlook}
In this work, we investigated the reliability of accelerated SH scheme, which relies on the scaling of SOCs and subsequent extrapolation to determine long-range time constants.
Using both \texttt{Fit1} and \texttt{Fit2} methods, we demonstrated that small trajectory ensembles are insufficient for providing reliable results.
The \texttt{Fit1} approach relies on a fixed theoretical scaling of $\alpha^{-2}$, whereas the \texttt{Fit2} approach utilizes a log--log transformation to fit the scaling exponent as a free parameter.
The latter is preferred as it does not restrict the exponent to $-$2, a constraint strictly applicable only to two-state systems, although our fits yielded values close to this theoretical value.
However, the additional fitting parameter increases the bootstrapping errors, an effect that can be mitigated by using a sufficiently large trajectory ensemble.
We recommend employing a multi-curve fitting strategy, which ensures stable extrapolation and yields physically reasonable slopes.
For the specific case of silaethylene, our analysis indicates that at least 100 trajectories per ensemble are necessary to achieve converged time constants and an even larger number of trajectories is required to reduce the bootstrapping errors.

Furthermore, we explored the integration of ML to enhance reliability of accelerated SH scheme.
We trained an MS-ANI model for the singlet PESs and separate ANI model for the triplet state.
The RPR procedure, combined with the gradient difference descriptor and a phase-correction algorithm, was used to fit NACs.
This approach was extended to learning of SOCs directly from Cartesian coordinates, translated to center of mass and rotated using tensor of inertia.
All models achieved satisfactory performance in preliminary static tests.

Our assessment of various ML enhancement opportunities revealed that increasing the number of trajectory ensembles ($N$) does not significantly improve the extrapolation scheme compared to increasing the number of trajectories ($n$) within each ensemble.
If the scaling factor is too small, the triplet population growth becomes difficult to fit; conversely, extending the simulation time to compensate for slow growth might lead to instabilities as the ML model extrapolates beyond its training data range.
Furthermore, large scaling factors might amplify the inaccuracies in the SOCs as shown for scaling factor 35, negatively influencing the resulting populations.

In conclusion, the most promising application of ML in this context is to expand the size of an ensembles, provided the ML models possess sufficient accuracy.
While the $S_1\rightarrow S_0$ relaxation is in excellent agreement and ML-predicted populations generally fall within the 95\% confidence interval of the MR-CISD reference, the observed discrepancies in the fitted time constants suggest that special care must be taken when assembling training sets.
Future work will focus on the implementation of AL procedures to refine these models and improve the overall robustness of ML-accelerated nonadiabatic dynamics.

\subsection{Data and code availability}
The data underlying this study, the Jupyter notebook used for the analysis, and any additional scripts and code are available from FigShare: \url{https://doi.org/10.6084/m9.figshare.33071360}.
The machine learning part of this work was done using open-source MLatom under the Apache 2.0 License in \url{https://github.com/dralgroup/mlatom}.

\subsection{Authors contributions}
J.M. generated the MR-CISD reference data, trained the models, performed final calculations and their analysis, and wrote the original manuscript.
M.K.S. implemented MLatom/Newton-X interface, generated the MR-CISD reference data, contributed to the result analysis and interpretation.
P.O.D. contributed to the result analysis and interpretation, co-supervised research.
J.P. designed and conceived the project, contributed to the result analysis and interpretation, co-supervised research, and secured funding.
All authors discussed the results and revised the manuscript.

\begin{acknowledgement}
The work of the Czech team has been supported by the \textit{Czech Science Foundation} Grant \mbox{23-06364S}, the \textit{Charles University} (project GAUK 6224) and
by the Advanced Multiscale Materials for Key Enabling Technologies project of the \textit{Ministry of Education, Youth, and Sports of the Czech Republic}, project No. CZ.02.01.01/00/22\_008/0004558, co-funded by the European Union.
We also highly appreciate the generous admission to computing facilities owned by parties and projects contributing to the National Grid Infrastructure MetaCentrum provided under the program ``Projects of Large Infrastructure for Research, Development, and Innovations ''(no. LM2010005) and computer time provided by the IT4I supercomputing center (Project ID:90254) supported by the Ministry of Education, Youth and Sports.
P.O.D. acknowledges funding by the projects for International Senior Scientists (Project No.: W2531013) and for Outstanding Youth Scholars (Overseas, 2021) of the National Natural Science Foundation of China, via the Lab project of the State Key Laboratory of Physical Chemistry of Solid Surfaces.
\end{acknowledgement}

\bibliography{ALL}

@Article{PhysNet2019,
  author  = {Unke, O. T. and Meuwly, M.},
  journal = {J. Chem. Theory Comput.},
  title   = {PhysNet: A Neural Network for Predicting Energies, Forces, Dipole Moments, and Partial Charges},
  year    = {2019},
  issn    = {1549-9626 (Electronic)
1549-9618 (Linking)},
  number  = {6},
  pages   = {3678–-3693},
  volume  = {15},
  doi     = {10.1021/acs.jctc.9b00181},
  type    = {Journal Article},
  url     = {http://www.ncbi.nlm.nih.gov/pubmed/31042390},
}

@Misc{Columbus7.2,
  author  = {Lischka, Hans and Shepard, Ron and Shavitt, Isaiah and Pitzer, Russel M. and M\"{u}ller, Thomas and Szalay, Péter G. and Brozell, Scott R. and Kedziora, Gary and Stahlberg, Eric A. and Harrison, Robert J. and Nieplocha, Jaroslaw and Minkoff, Michael and Barbatti, Mario and Schuurmann, Michael and Guan, Yafu and Yarkony, David R. and Matsika, Spiridoula and Plasser, Felix and Beck, Eric. V. and Blaudeau, Jean-Philippe and Ruckenbauer, Matthias and Sellner, Bernhard and Szymczak, Jaroslaw J. and Spada, Rene F. K. and Das, Anita and Belcher, Lachlan T. and Nieman, Reed},
  title   = {COLUMBUS, an Ab Initio Electronic Structure Program, Release 7.2},
  year    = {2022},
  comment = {Columbus 7.2},
}

@Article{Li2024a,
  author    = {Li, Shuai and Xie, Bin-Bin and Yin, Bo-Wen and Liu, Lihong and Shen, Lin and Fang, Wei-Hai},
  journal   = {J. Phys. Chem. A},
  title     = {Construction of Highly Accurate Machine Learning Potential Energy Surfaces for Excited-State Dynamics Simulations Based on Low-Level Data Sets},
  year      = {2024},
  issn      = {1520-5215},
  month     = jul,
  number    = {28},
  pages     = {5516--5524},
  volume    = {128},
  doi       = {10.1021/acs.jpca.4c02028},
  fjournal  = {The Journal of Physical Chemistry A},
  publisher = {American Chemical Society (ACS)},
}

@Article{Chen2018,
  author    = {Wen-Kai Chen and Xiang-Yang Liu and Wei-Hai Fang and Pavlo O. Dral and Ganglong Cui},
  journal   = {J. Phys. Chem. Lett.},
  title     = {Deep Learning for Nonadiabatic Excited-State Dynamics},
  year      = {2018},
  month     = {nov},
  number    = {23},
  pages     = {6702--6708},
  volume    = {9},
  doi       = {10.1021/acs.jpclett.8b03026},
  fjournal  = {The Journal of Physical Chemistry Letters},
  publisher = {American Chemical Society ({ACS})},
}

@Article{Richter2011,
  author    = {Richter, Martin and Marquetand, Philipp and Gonz\'{a}lez-V\'{a}zquez, Jesús and Sola, Ignacio and Gonz\'{a}lez, Leticia},
  journal   = {J. Chem. Theory Comput.},
  title     = {SHARC: ab Initio Molecular Dynamics with Surface Hopping in the Adiabatic Representation Including Arbitrary Couplings},
  year      = {2011},
  issn      = {1549-9626},
  month     = mar,
  number    = {5},
  pages     = {1253--1258},
  volume    = {7},
  comment   = {old SHARC},
  doi       = {10.1021/ct1007394},
  fjournal  = {Journal of Chemical Theory and Computation},
  publisher = {American Chemical Society (ACS)},
}

@Article{Chen2019,
  author    = {Chen, Wen-Kai and Fang, Wei-Hai and Cui, Ganglong},
  journal   = {J. Phys. Chem. Lett.},
  title     = {Integrating Machine Learning with the Multilayer Energy-Based Fragment Method for Excited States of Large Systems},
  year      = {2019},
  issn      = {1948-7185},
  month     = dec,
  number    = {24},
  pages     = {7836--7841},
  volume    = {10},
  doi       = {10.1021/acs.jpclett.9b03113},
  fjournal  = {The Journal of Physical Chemistry Letters},
  publisher = {American Chemical Society (ACS)},
}

@Article{Dral2017,
  author    = {Pavlo O. Dral and Alec Owens and Sergei N. Yurchenko and Walter Thiel},
  journal   = {J. Chem. Phys.},
  title     = {Structure-Based Sampling and Self-Correcting Machine Learning for Accurate Calculations of Potential Energy Surfaces and Vibrational Levels},
  year      = {2017},
  month     = {jun},
  number    = {24},
  pages     = {244108},
  volume    = {146},
  doi       = {10.1063/1.4989536},
  fjournal  = {The Journal of Chemical Physics},
  publisher = {{AIP} Publishing},
}

@Article{Mai2015a,
  author    = {Mai, Sebastian and Marquetand, Philipp and Gonz\'{a}lez, Leticia},
  journal   = {Int. J. Quantum Chem.},
  title     = {A General Method to Describe Intersystem Crossing Dynamics in Trajectory Surface Hopping},
  year      = {2015},
  issn      = {1097-461X},
  month     = mar,
  number    = {18},
  pages     = {1215--1231},
  volume    = {115},
  comment   = {propagator matrix, optimal representation for SH is discussed},
  doi       = {10.1002/qua.24891},
  fjournal  = {International Journal of Quantum Chemistry},
  publisher = {Wiley},
}

@Article{Sit2024,
  author    = {Sit, Mahesh K. and Das, Subhasish and Samanta, Kousik},
  journal   = {J. Phys. Chem. A},
  title     = {Machine Learning-Assisted Mixed Quantum-Classical Dynamics without Explicit Nonadiabatic Coupling: Application to the Photodissociation of Peroxynitric Acid},
  year      = {2024},
  issn      = {1520-5215},
  month     = sep,
  number    = {38},
  pages     = {8244--8253},
  volume    = {128},
  doi       = {10.1021/acs.jpca.4c02876},
  fjournal  = {The Journal of Physical Chemistry A},
  publisher = {American Chemical Society (ACS)},
}

@Article{WasifBaig2025,
  author    = {Wasif Baig, Mirza and Pederzoli, Marek and Kývala, Mojmír and Pittner, Jiří},
  journal   = {J. Comput. Chem.},
  title     = {Quantum Chemical and Trajectory Surface Hopping Molecular Dynamics Study of Iodine‐Based BODIPY Photosensitizer},
  year      = {2025},
  issn      = {1096-987X},
  month     = mar,
  number    = {7},
  pages     = {e70026},
  volume    = {46},
  doi       = {10.1002/jcc.70026},
  fjournal  = {Journal of Computational Chemistry},
  publisher = {Wiley},
}

@Article{Mukherjee2022a,
  author    = {Mukherjee, Saikat and Pinheiro, Max and Demoulin, Baptiste and Barbatti, Mario},
  journal   = {Philos. Trans. R. Soc. Math. Phys. Eng. Sci.},
  title     = {Simulations of molecular photodynamics in long timescales},
  year      = {2022},
  issn      = {1471-2962},
  month     = mar,
  number    = {2223},
  pages     = {20200382},
  volume    = {380},
  comment   = {bad ML PES after 600 fs},
  doi       = {10.1098/rsta.2020.0382},
  fjournal  = {Philosophical Transactions of the Royal Society A: Mathematical, Physical and Engineering Sciences},
  publisher = {The Royal Society},
}

@Article{Sit2023,
  author    = {Mahesh K. Sit and Subhasish Das and Kousik Samanta},
  journal   = {J. Phys. Chem. A},
  title     = {Semiclassical Dynamics on Machine-Learned Coupled Multireference Potential Energy Surfaces: Application to the Photodissociation of the Simplest Criegee Intermediate},
  year      = {2023},
  month     = {mar},
  number    = {10},
  pages     = {2376--2387},
  volume    = {127},
  doi       = {10.1021/acs.jpca.2c07229},
  fjournal  = {The Journal of Physical Chemistry A},
  publisher = {American Chemical Society ({ACS})},
}

@Article{Pitonak2005,
  author    = {Pitonak, M. and Lischka, H.},
  journal   = {Mol. Phys.},
  title     = {Excited-state potential energy surfaces of silaethylene: a MRCI investigation},
  year      = {2005},
  issn      = {1362-3028},
  month     = mar,
  number    = {6–8},
  pages     = {855--862},
  volume    = {103},
  doi       = {10.1080/00268970412331333573},
  fjournal  = {Molecular Physics},
  publisher = {Informa UK Limited},
}

@Article{Akimov2021,
  author    = {Akimov, Alexey V.},
  journal   = {J. Phys. Chem. Lett.},
  title     = {Extending the Time Scales of Nonadiabatic Molecular Dynamics via Machine Learning in the Time Domain},
  year      = {2021},
  issn      = {1948-7185},
  month     = dec,
  number    = {50},
  pages     = {12119--12128},
  volume    = {12},
  doi       = {10.1021/acs.jpclett.1c03823},
  fjournal  = {The Journal of Physical Chemistry Letters},
  publisher = {American Chemical Society (ACS)},
}

@Article{CrespoOtero2018,
  author    = {Crespo-Otero, Rachel and Barbatti, Mario},
  journal   = {Chem. Rev.},
  title     = {Recent Advances and Perspectives on Nonadiabatic Mixed Quantum–Classical Dynamics},
  year      = {2018},
  issn      = {1520-6890},
  month     = may,
  number    = {15},
  pages     = {7026--7068},
  volume    = {118},
  comment   = {single-reference methods only for dynamics which does not involve decay to the ground state},
  doi       = {10.1021/acs.chemrev.7b00577},
  fjournal  = {Chemical Reviews},
  publisher = {American Chemical Society (ACS)},
}

@Article{Ge2024,
  author    = {Ge, Fuchun and Wang, Ran and Qu, Chen and Zheng, Peikun and Nandi, Apurba and Conte, Riccardo and Houston, Paul L. and Bowman, Joel M. and Dral, Pavlo O.},
  journal   = {J. Phys. Chem. Lett.},
  title     = {Tell Machine Learning Potentials What They Are Needed For: Simulation-Oriented Training Exemplified for Glycine},
  year      = {2024},
  issn      = {1948-7185},
  month     = apr,
  number    = {16},
  pages     = {4451--4460},
  volume    = {15},
  doi       = {10.1021/acs.jpclett.4c00746},
  fjournal  = {The Journal of Physical Chemistry Letters},
  publisher = {American Chemical Society (ACS)},
}

@Article{Pinheiro2021,
  author    = {Max {Pinheiro Jr} and Fuchun Ge and Nicolas Ferr{\'{e}} and Pavlo O. Dral and Mario Barbatti},
  journal   = {Chem. Sci.},
  title     = {Choosing the Right Molecular Machine Learning Potential},
  year      = {2021},
  number    = {43},
  pages     = {14396--14413},
  volume    = {12},
  doi       = {10.1039/d1sc03564a},
  fjournal  = {Chemical Science},
  publisher = {Royal Society of Chemistry ({RSC})},
}

@Article{Bartok2013,
  author    = {Bartók, Albert P. and Kondor, Risi and Cs\'{a}nyi, G\'{a}bor},
  journal   = {Phys. Rev. B},
  title     = {On Representing Chemical Environments},
  year      = {2013},
  issn      = {1550-235X},
  month     = may,
  number    = {18},
  pages     = {184115},
  volume    = {87},
  comment   = {SOAP},
  doi       = {10.1103/physrevb.87.184115},
  fjournal  = {Physical Review B},
  publisher = {American Physical Society (APS)},
}

@Article{Zechmann2006,
  author    = {Gunther Zechmann and Mario Barbatti and Hans Lischka and Jiří Pittner and Vlasta Bona{\v{c}}i{\'{c}}-Kouteck{\'{y}}},
  journal   = {Chem. Phys. Lett.},
  title     = {Multiple pathways in the photodynamics of a polar $\pi$-bond: A case study of silaethylene},
  year      = {2006},
  month     = {feb},
  number    = {4-6},
  pages     = {377--382},
  volume    = {418},
  doi       = {10.1016/j.cplett.2005.11.015},
  fjournal  = {Chemical Physics Letters},
  publisher = {Elsevier {BV}},
}

@Article{Wilkins2019,
  author    = {Wilkins, David M. and Grisafi, Andrea and Yang, Yang and Lao, Ka Un and DiStasio, Robert A. and Ceriotti, Michele},
  journal   = {Proc. Natl. Acad. Sci.},
  title     = {Accurate molecular polarizabilities with coupled cluster theory and machine learning},
  year      = {2019},
  issn      = {1091-6490},
  month     = feb,
  number    = {9},
  pages     = {3401--3406},
  volume    = {116},
  comment   = {SA-GPR tensor learning, generalizes reference frame approach},
  doi       = {10.1073/pnas.1816132116},
  fjournal  = {Proceedings of the National Academy of Sciences},
  publisher = {Proceedings of the National Academy of Sciences},
}

@Article{Noe2020,
  author    = {Frank No{\'{e}} and Alexandre Tkatchenko and Klaus-Robert M\"{u}ller and Cecilia Clementi},
  journal   = {Annu. Rev. Phys. Chem.},
  title     = {Machine Learning for Molecular Simulation},
  year      = {2020},
  month     = {apr},
  number    = {1},
  pages     = {361--390},
  volume    = {71},
  doi       = {10.1146/annurev-physchem-042018-052331},
  fjournal  = {Annual Review of Physical Chemistry},
  publisher = {Annual Reviews},
}

@Article{Tully1990,
  author    = {Tully, John C.},
  journal   = {J. Chem. Phys.},
  title     = {Molecular Dynamics with Electronic Transitions},
  year      = {1990},
  issn      = {1089-7690},
  month     = jul,
  number    = {2},
  pages     = {1061--1071},
  volume    = {93},
  comment   = {overcoherence},
  doi       = {10.1063/1.459170},
  fjournal  = {The Journal of Chemical Physics},
  publisher = {AIP Publishing},
  ranking   = {rank5},
}

@Article{Jumper2021,
  author    = {John Jumper and Richard Evans and Alexander Pritzel and Tim Green and Michael Figurnov and Olaf Ronneberger and Kathryn Tunyasuvunakool and Russ Bates and Augustin {\v{Z}}{\'{\i}}dek and Anna Potapenko and Alex Bridgland and Clemens Meyer and Simon A. A. Kohl and Andrew J. Ballard and Andrew Cowie and Bernardino Romera-Paredes and Stanislav Nikolov and Rishub Jain and Jonas Adler and Trevor Back and Stig Petersen and David Reiman and Ellen Clancy and Michal Zielinski and Martin Steinegger and Michalina Pacholska and Tamas Berghammer and Sebastian Bodenstein and David Silver and Oriol Vinyals and Andrew W. Senior and Koray Kavukcuoglu and Pushmeet Kohli and Demis Hassabis},
  journal   = {Nat},
  title     = {Highly accurate protein structure prediction with {AlphaFold}},
  year      = {2021},
  month     = {jul},
  number    = {7873},
  pages     = {583--589},
  volume    = {596},
  comment   = {NP 2024 Chemistry, AlphaFold},
  doi       = {10.1038/s41586-021-03819-2},
  fjournal  = {Nature},
  publisher = {Springer Science and Business Media {LLC}},
}

@Article{Mukherjee2022,
  author   = {Mukherjee, Saikat and Barbatti, Mario},
  journal  = {J. Chem. Theory Comput.},
  title    = {A Hessian-Free Method to Prevent Zero-Point Energy Leakage in Classical Trajectories},
  year     = {2022},
  number   = {7},
  pages    = {4109-4116},
  volume   = {18},
  doi      = {10.1021/acs.jctc.2c00216},
  eprint   = {https://doi.org/10.1021/acs.jctc.2c00216},
  fjournal = {Journal of Chemical Theory and Computation},
  url      = {https://doi.org/10.1021/acs.jctc.2c00216},
}

@Article{Mausenberger2024a,
  author     = {Mausenberger, Sascha and M\"{u}ller, Carolin and Tkatchenko, Alexandre and Marquetand, Philipp and Gonz\'{a}lez, Leticia and Westermayr, Julia},
  journal    = {Chem. Sci.},
  title      = {{SPAINN}: Equivariant Message Passing for Excited-State Nonadiabatic Molecular Dynamics},
  year       = {2024},
  issn       = {2041-6539},
  month      = oct,
  number     = {38},
  pages      = {15880--15890},
  volume     = {15},
  doi        = {10.1039/D4SC04164J},
  fjournal   = {Chemical Science},
  language   = {en},
  publisher  = {The Royal Society of Chemistry},
  shorttitle = {{SPAINN}},
  url        = {https://pubs.rsc.org/en/content/articlelanding/2024/sc/d4sc04164j},
  urldate    = {2024-11-29},
}

@Article{Martinka2024,
  author    = {Martinka, Jakub and Pederzoli, Marek and Barbatti, Mario and Dral, Pavlo O. and Pittner, Jiří},
  journal   = {J. Chem. Phys.},
  title     = {A Simple Approach to Rotationally Invariant Machine Learning of a Vector Quantity},
  year      = {2024},
  issn      = {1089-7690},
  month     = nov,
  pages     = {174104},
  number    = {17},
  volume    = {161},
  doi       = {10.1063/5.0230176},
  fjournal  = {The Journal of Chemical Physics},
  publisher = {AIP Publishing},
}

@Article{Dral2021,
  author    = {Pavlo O. Dral and Mario Barbatti},
  journal   = {Nat. Rev. Chem.},
  title     = {Molecular Excited States Through a Machine Learning Lens},
  year      = {2021},
  month     = {may},
  number    = {6},
  pages     = {388--405},
  volume    = {5},
  doi       = {10.1038/s41570-021-00278-1},
  fjournal  = {Nature Reviews Chemistry},
  publisher = {Springer Science and Business Media {LLC}},
}

@Article{Grisafi2018,
  author    = {Andrea Grisafi and David M. Wilkins and G{\'{a}}bor Cs{\'{a}}nyi and Michele Ceriotti},
  journal   = {Phys. Rev. Lett.},
  title     = {Symmetry-Adapted Machine Learning for Tensorial Properties of Atomistic Systems},
  year      = {2018},
  month     = {jan},
  number    = {3},
  pages     = {036002},
  volume    = {120},
  comment   = {SA-GPR tensor learning, generalizes reference frame approach},
  doi       = {10.1103/physrevlett.120.036002},
  fjournal  = {Physical Review Letters},
  publisher = {American Physical Society ({APS})},
}

@Article{Suchan2020,
  author    = {Suchan, Jiří and Janoš, Jiří and Slavíček, Petr},
  journal   = {J. Chem. Theory Comput.},
  title     = {Pragmatic Approach to Photodynamics: Mixed Landau–Zener Surface Hopping with Intersystem Crossing},
  year      = {2020},
  issn      = {1549-9626},
  month     = jul,
  number    = {9},
  pages     = {5809--5820},
  volume    = {16},
  comment   = {performance seems to be satisfactory under certain conditions, with superior stability over other methods},
  doi       = {10.1021/acs.jctc.0c00512},
  fjournal  = {Journal of Chemical Theory and Computation},
  publisher = {American Chemical Society (ACS)},
}

@Article{Westermayr2020c,
  author    = {Westermayr, Julia and Marquetand, Philipp},
  journal   = {Chem. Rev.},
  title     = {Machine Learning for Electronically Excited States of Molecules},
  year      = {2020},
  issn      = {1520-6890},
  month     = nov,
  number    = {16},
  pages     = {9873--9926},
  volume    = {121},
  doi       = {10.1021/acs.chemrev.0c00749},
  fjournal  = {Chemical Reviews},
  publisher = {American Chemical Society (ACS)},
}

@Article{MLatom2024,
  author    = {Dral, Pavlo O. and Ge, Fuchun and Hou, Yi-Fan and Zheng, Peikun and Chen, Yuxinxin and Barbatti, Mario and Isayev, Olexandr and Wang, Cheng and Xue, Bao-Xin and Pinheiro Jr, Max and Su, Yuming and Dai, Yiheng and Chen, Yangtao and Zhang, Lina and Zhang, Shuang and Ullah, Arif and Zhang, Quanhao and Ou, Yanchi},
  journal   = {J. Chem. Theory Comput.},
  title     = {MLatom 3: A Platform for Machine Learning-Enhanced Computational Chemistry Simulations and Workflows},
  year      = {2024},
  issn      = {1549-9626},
  month     = feb,
  number    = {3},
  pages     = {1193-1213},
  volume    = {20},
  comment   = {MLatom 3},
  day       = {13},
  doi       = {10.1021/acs.jctc.3c01203},
  fjournal  = {Journal of Chemical Theory and Computation},
  publisher = {American Chemical Society (ACS)},
}

@Article{Martyka2025,
  author        = {Martyka, Mikołaj and Zhang, Lina and Ge, Fuchun and Hou, Yi-Fan and Jankowska, Joanna and Barbatti, Mario and Dral, Pavlo O.},
  journal       = {npj Comput. Mater.},
  title         = {Charting Electronic-State Manifolds Across Molecules with Multi-State Learning and Gap-Driven Dynamics via Efficient and Robust Active Learning},
  year          = {2025},
  issn          = {2057-3960},
  month         = may,
  number        = {1},
  pages         = {132},
  volume        = {11},
  comment       = {MS-ANI},
  comment-jakub = {MSANI},
  doi           = {10.1038/s41524-025-01636-z},
  fjournal      = {npj Computational Materials},
  publisher     = {Springer Science and Business Media LLC},
}

@Article{Li2021,
  author    = {Li, Jingbai and Stein, Rachel and Adrion, Daniel M. and Lopez, Steven A.},
  journal   = {J. Am. Chem. Soc.},
  title     = {Machine-Learning Photodynamics Simulations Uncover the Role of Substituent Effects on the Photochemical Formation of Cubanes},
  year      = {2021},
  issn      = {1520-5126},
  month     = nov,
  number    = {48},
  pages     = {20166--20175},
  volume    = {143},
  comment   = {PyRAI2MD, Zhu-Nakamura},
  doi       = {10.1021/jacs.1c07725},
  fjournal  = {Journal of the American Chemical Society},
  publisher = {American Chemical Society (ACS)},
}

@Article{Tiefenbacher2025,
  author    = {Tiefenbacher, Maximilian X. and Bachmair, Brigitta and Chen, Cheng Giuseppe and Westermayr, Julia and Marquetand, Philipp and Dietschreit, Johannes C. B. and González, Leticia},
  journal   = {Digit. Discov.},
  title     = {Excited-state nonadiabatic dynamics in explicit solvent using machine learned interatomic potentials},
  year      = {2025},
  issn      = {2635-098X},
  number    = {6},
  pages     = {1478--1491},
  volume    = {4},
  comment   = {Data split},
  doi       = {10.1039/d5dd00044k},
  fjournal  = {Digital Discovery},
  publisher = {Royal Society of Chemistry (RSC)},
}

@Article{Granucci2012,
  author    = {Granucci, Giovanni and Persico, Maurizio and Spighi, Gloria},
  journal   = {J. Chem. Phys.},
  title     = {Surface hopping trajectory simulations with spin-orbit and dynamical couplings},
  year      = {2012},
  issn      = {1089-7690},
  month     = jul,
  number    = {22},
  pages     = {22A501},
  volume    = {137},
  comment   = {optimal representation for SH discussed, multiplet components in ISC FSSH has to be treated separately, decoherence should not be used when using SOCs},
  doi       = {10.1063/1.4707737},
  fjournal  = {The Journal of Chemical Physics},
  publisher = {AIP Publishing},
}

@Article{Behler2007,
  author    = {Behler, Jörg and Parrinello, Michele},
  journal   = {Phys. Rev. Lett.},
  title     = {Generalized Neural-Network Representation of High-Dimensional Potential-Energy Surfaces},
  year      = {2007},
  issn      = {1079-7114},
  month     = apr,
  number    = {14},
  pages     = {146401},
  volume    = {98},
  comment   = {Symmetry functions; if forces in the data set MLP accuracy improves},
  doi       = {10.1103/physrevlett.98.146401},
  fjournal  = {Physical Review Letters},
  publisher = {American Physical Society (APS)},
}

@Article{Lingerfelt2016,
  author    = {Lingerfelt, David B. and Williams-Young, David B. and Petrone, Alessio and Li, Xiaosong},
  journal   = {J. Chem. Theory Comput.},
  title     = {Direct ab Initio (Meta-)Surface-Hopping Dynamics},
  year      = {2016},
  issn      = {1549-9626},
  month     = feb,
  number    = {3},
  pages     = {935--945},
  volume    = {12},
  comment   = {Scaling SOCs},
  doi       = {10.1021/acs.jctc.5b00697},
  fjournal  = {Journal of Chemical Theory and Computation},
  publisher = {American Chemical Society (ACS)},
}

@Article{Nijamudheen2017,
  author    = {Nijamudheen, A. and Akimov, Alexey V.},
  journal   = {The Journal of Physical Chemistry C},
  title     = {Excited-State Dynamics in Two-Dimensional Heterostructures: SiR/TiO$_2$ and GeR/TiO$_2$ (R$=$H, Me) as Promising Photocatalysts},
  year      = {2017},
  issn      = {1932-7455},
  month     = mar,
  number    = {12},
  pages     = {6520--6532},
  volume    = {121},
  comment   = {Scaling SOCs},
  doi       = {10.1021/acs.jpcc.7b00545},
  publisher = {American Chemical Society (ACS)},
}

@Article{Lischka2020,
  author    = {Lischka, Hans and Shepard, Ron and M\"{u}ller, Thomas and Szalay, Péter G. and Pitzer, Russell M. and Aquino, Adelia J. A. and Araújo do Nascimento, Mayzza M. and Barbatti, Mario and Belcher, Lachlan T. and Blaudeau, Jean-Philippe and Borges, Itamar and Brozell, Scott R. and Carter, Emily A. and Das, Anita and Gidofalvi, Gergely and Gonz\'{a}lez, Leticia and Hase, William L. and Kedziora, Gary and Kertesz, Miklos and Kossoski, F\'{a}bris and Machado, Francisco B. C. and Matsika, Spiridoula and do Monte, Silmar A. and Nachtigallov\'{a}, Dana and Nieman, Reed and Oppel, Markus and Parish, Carol A. and Plasser, Felix and Spada, Rene F. K. and Stahlberg, Eric A. and Ventura, Elizete and Yarkony, David R. and Zhang, Zhiyong},
  journal   = {J. Chem. Phys.},
  title     = {The Generality of the GUGA MRCI Approach in COLUMBUS for Treating Complex Quantum Chemistry},
  year      = {2020},
  issn      = {1089-7690},
  month     = apr,
  number    = {13},
  pages     = {134110},
  volume    = {152},
  comment   = {COLUMBUS},
  doi       = {10.1063/1.5144267},
  fjournal  = {The Journal of Chemical Physics},
  publisher = {AIP Publishing},
}

@Article{Li2021a,
  author    = {Li, Jingbai and Reiser, Patrick and Boswell, Benjamin R. and Eberhard, André and Burns, Noah Z. and Friederich, Pascal and Lopez, Steven A.},
  journal   = {Chem. Sci.},
  title     = {Automatic Discovery of Photoisomerization Mechanisms with Nanosecond Machine Learning Photodynamics Simulations},
  year      = {2021},
  issn      = {2041-6539},
  number    = {14},
  pages     = {5302--5314},
  volume    = {12},
  comment   = {PyRAI2MD, Zhu-Nakamura},
  doi       = {10.1039/d0sc05610c},
  fjournal  = {Chemical Science},
  publisher = {Royal Society of Chemistry (RSC)},
}

@Article{Heller2021,
  author    = {Heller, Eric R. and Joswig, Jan-Ole and Seifert, Gotthard},
  journal   = {Theor. Chem. Acc.},
  title     = {Exploring the effects of quantum decoherence on the excited-state dynamics of molecular systems},
  year      = {2021},
  issn      = {1432-2234},
  month     = apr,
  number    = {4},
  volume    = {140},
  comment   = {decoherence should not be used when using SOCs},
  doi       = {10.1007/s00214-021-02741-0},
  fjournal  = {Theoretical Chemistry Accounts},
  publisher = {Springer Science and Business Media LLC},
}

@Article{HammesSchiffer1994,
  author    = {Hammes-Schiffer, Sharon and Tully, John C.},
  journal   = {J. Chem. Phys.},
  title     = {Proton Transfer in Solution: Molecular Dynamics with Quantum Transitions},
  year      = {1994},
  issn      = {1089-7690},
  month     = sep,
  number    = {6},
  pages     = {4657--4667},
  volume    = {101},
  comment   = {TDCs from WF overlap},
  doi       = {10.1063/1.467455},
  fjournal  = {The Journal of Chemical Physics},
  publisher = {AIP Publishing},
  ranking   = {rank5},
}

@Article{Bispo2025,
  author    = {Bispo, Matheus de Oliveira and Souza Mattos, Rafael and Pinheiro, Max and Garain, Bidhan Chandra and Dral, Pavlo O. and Barbatti, Mario},
  journal   = {J. Chem. Theory Comput.},
  title     = {MELTS: Fully Automated Active Learning for Fewest-Switches Surface Hopping Dynamics},
  year      = {2025},
  issn      = {1549-9626},
  month     = nov,
  number    = {22},
  pages     = {11390--11400},
  volume    = {21},
  comment   = {MELTS},
  doi       = {10.1021/acs.jctc.5c01454},
  fjournal  = {Journal of Chemical Theory and Computation},
  publisher = {American Chemical Society (ACS)},
}

@Article{Ramos2025,
  author    = {Ramos, Mayk Caldas and Collison, Christopher J. and White, Andrew D.},
  journal   = {Chem. Sci.},
  title     = {A review of large language models and autonomous agents in chemistry},
  year      = {2025},
  issn      = {2041-6539},
  number    = {6},
  pages     = {2514--2572},
  volume    = {16},
  doi       = {10.1039/d4sc03921a},
  fjournal  = {Chemical Science},
  publisher = {Royal Society of Chemistry (RSC)},
}

@Article{Bartok2010,
  author    = {Bartók, Albert P. and Payne, Mike C. and Kondor, Risi and Csányi, Gábor},
  journal   = {Phys. Rev. Lett.},
  title     = {Gaussian Approximation Potentials: The Accuracy of Quantum Mechanics, without the Electrons},
  year      = {2010},
  issn      = {1079-7114},
  month     = apr,
  number    = {13},
  pages     = {136403},
  volume    = {104},
  comment   = {GAP},
  doi       = {10.1103/physrevlett.104.136403},
  fjournal  = {Physical Review Letters},
  publisher = {American Physical Society (APS)},
}

@Article{Marcus1956,
  author    = {Marcus, R. A.},
  journal   = {J. Chem. Phys.},
  title     = {On the Theory of Oxidation-Reduction Reactions Involving Electron Transfer. I},
  year      = {1956},
  issn      = {1089-7690},
  month     = may,
  number    = {5},
  pages     = {966--978},
  volume    = {24},
  comment   = {Marcus theory},
  doi       = {10.1063/1.1742723},
  fjournal  = {The Journal of Chemical Physics},
  publisher = {AIP Publishing},
}

@Article{Marcus1993,
  author    = {Marcus, Rudolph A.},
  fjournal  = {Angew. Chem. Int. Ed. Engl.},
  title     = {Electron Transfer Reactions in Chemistry: Theory and Experiment (Nobel Lecture)},
  year      = {1993},
  issn      = {0570-0833},
  month     = aug,
  number    = {8},
  pages     = {1111--1121},
  volume    = {32},
  comment   = {Marcus theory},
  doi       = {10.1002/anie.199311113},
  fjournal  = {Angewandte Chemie International Edition in English},
  publisher = {Wiley},
}

@Article{Pederzoli2017,
  author    = {Marek Pederzoli and Ji{\v{r}}{\'{\i}} Pittner},
  journal   = {J. Chem. Phys.},
  title     = {A new approach to molecular dynamics with non-adiabatic and spin-orbit effects with applications to {QM}/{MM} simulations of thiophene and selenophene},
  year      = {2017},
  month     = mar,
  number    = {11},
  pages     = {114101},
  volume    = {146},
  comment   = {Newton-X ISC},
  doi       = {10.1063/1.4978289},
  fjournal  = {The Journal of Chemical Physics},
  publisher = {{AIP} Publishing},
}

@Article{Pederzoli2019,
  author    = {Pederzoli, Marek and Wasif Baig, Mirza and Kývala, Mojmír and Pittner, Jiří and Cwiklik, Lukasz},
  journal   = {J. Chem. Theory Comput.},
  title     = {Photophysics of BODIPY-Based Photosensitizer for Photodynamic Therapy: Surface Hopping and Classical Molecular Dynamics},
  year      = {2019},
  issn      = {1549-9626},
  month     = aug,
  number    = {9},
  pages     = {5046--5057},
  volume    = {15},
  doi       = {10.1021/acs.jctc.9b00533},
  fjournal  = {Journal of Chemical Theory and Computation},
  publisher = {American Chemical Society (ACS)},
}

@Article{Westermayr2019,
  author    = {Julia Westermayr and Michael Gastegger and Maximilian F. S. J. Menger and Sebastian Mai and Leticia Gonz{\'{a}}lez and Philipp Marquetand},
  journal   = {Chem. Sci.},
  title     = {Machine Learning Enables Long Time Scale Molecular Photodynamics Simulations},
  year      = {2019},
  number    = {35},
  pages     = {8100--8107},
  volume    = {10},
  doi       = {10.1039/c9sc01742a},
  fjournal  = {Chemical Science},
  publisher = {Royal Society of Chemistry ({RSC})},
}

@Article{Dral2020,
  author    = {Pavlo O. Dral},
  journal   = {J. Phys. Chem. Lett.},
  title     = {Quantum Chemistry in the Age of Machine Learning},
  year      = {2020},
  month     = {mar},
  number    = {6},
  pages     = {2336--2347},
  volume    = {11},
  doi       = {10.1021/acs.jpclett.9b03664},
  fjournal  = {The Journal of Physical Chemistry Letters},
  publisher = {American Chemical Society ({ACS})},
}

@Article{Westermayr2020b,
  author    = {Julia Westermayr and Michael Gastegger and Philipp Marquetand},
  journal   = {J. Phys. Chem. Lett.},
  title     = {Combining {SchNet} and {SHARC}: The {SchNarc} Machine Learning Approach for Excited-State Dynamics},
  year      = {2020},
  month     = {apr},
  number    = {10},
  pages     = {3828--3834},
  volume    = {11},
  comment   = {phase-less loss learning of NACs},
  doi       = {10.1021/acs.jpclett.0c00527},
  fjournal  = {The Journal of Physical Chemistry Letters},
  publisher = {American Chemical Society ({ACS})},
}

@Article{Dral2018,
  author    = {Pavlo O. Dral and Mario Barbatti and Walter Thiel},
  journal   = {J. Phys. Chem. Lett.},
  title     = {Nonadiabatic Excited-State Dynamics with Machine Learning},
  year      = {2018},
  month     = {sep},
  number    = {19},
  pages     = {5660--5663},
  volume    = {9},
  comment   = {reported difficulties in the ML-prediction of nonadiabatic couplings},
  doi       = {10.1021/acs.jpclett.8b02469},
  fjournal  = {The Journal of Physical Chemistry Letters},
  publisher = {American Chemical Society ({ACS})},
}

@Article{Mukherjee2024a,
  author    = {Mukherjee, Saikat and Lassmann, Yorick and Mattos, Rafael S. and Demoulin, Baptiste and Curchod, Basile F. E. and Barbatti, Mario},
  journal   = {J. Chem. Theory Comput.},
  title     = {Assessing Nonadiabatic Dynamics Methods in Long Timescales},
  year      = {2024},
  issn      = {1549-9626},
  month     = dec,
  number    = {1},
  pages     = {29--37},
  volume    = {21},
  doi       = {10.1021/acs.jctc.4c01349},
  fjournal  = {Journal of Chemical Theory and Computation},
  publisher = {American Chemical Society (ACS)},
}

@Article{Martinka2025,
  author    = {Martinka, Jakub and Zhang, Lina and Hou, Yi-Fan and Martyka, Mikołaj and Pittner, Jiří and Barbatti, Mario and Dral, Pavlo O.},
  journal   = {J. Phys. Chem. Lett.},
  title     = {A Descriptor Is All You Need: Accurate Machine Learning of Nonadiabatic Coupling Vectors},
  year      = {2025},
  issn      = {1948-7185},
  month     = nov,
  number    = {45},
  pages     = {11732--11744},
  volume    = {16},
  doi       = {10.1021/acs.jpclett.5c02810},
  fjournal  = {The Journal of Physical Chemistry Letters},
  publisher = {American Chemical Society (ACS)},
}

@Article{Akimov2017,
  author    = {Akimov, Alexey V.},
  journal   = {J. Phys. Chem. Lett.},
  title     = {Stochastic and Quasi-Stochastic Hamiltonians for Long-Time Nonadiabatic Molecular Dynamics},
  year      = {2017},
  issn      = {1948-7185},
  month     = oct,
  number    = {20},
  pages     = {5190--5195},
  volume    = {8},
  doi       = {10.1021/acs.jpclett.7b02185},
  fjournal  = {The Journal of Physical Chemistry Letters},
  publisher = {American Chemical Society (ACS)},
}

@Article{Li2020,
  author    = {Li, Jingbai and Reiser, Patrick and Eberhard, André and Friederich, Pascal and Lopez, Steven},
  title     = {Nanosecond Photodynamics Simulations of a Cis-Trans Isomerization Are Enabled by Machine Learning},
  year      = {2020},
  month     = oct,
  doi       = {10.26434/chemrxiv.13047863.v1},
  publisher = {American Chemical Society (ACS)},
}

@Article{Zou2025,
  author    = {Zou, Yunheng and Cheng, Austin H. and Aldossary, Abdulrahman and Bai, Jiaru and Leong, Shi Xuan and Campos-Gonzalez-Angulo, Jorge Arturo and Choi, Changhyeok and Ser, Cher Tian and Tom, Gary and Wang, Andrew and Zhang, Zijian and Yakavets, Ilya and Hao, Han and Crebolder, Chris and Bernales, Varinia and Aspuru-Guzik, Alán},
  journal   = {Matter},
  title     = {El Agente: An autonomous agent for quantum chemistry},
  year      = {2025},
  issn      = {2590-2385},
  month     = jul,
  number    = {7},
  pages     = {102263},
  volume    = {8},
  doi       = {10.1016/j.matt.2025.102263},
  publisher = {Elsevier BV},
}

@Article{Leong2025,
  author    = {Leong, Shi Xuan and Griesbach, Caleb E. and Zhang, Rui and Darvish, Kourosh and Zhao, Yuchi and Mandal, Abhijoy and Zou, Yunheng and Hao, Han and Bernales, Varinia and Aspuru-Guzik, Alán},
  journal   = {Nat. Rev. Chem.},
  title     = {Steering towards safe self-driving laboratories},
  year      = {2025},
  issn      = {2397-3358},
  month     = aug,
  number    = {10},
  pages     = {707--722},
  volume    = {9},
  comment   = {self-driving labs},
  doi       = {10.1038/s41570-025-00747-x},
  fjournal  = {Nature Reviews Chemistry},
  publisher = {Springer Science and Business Media LLC},
}

@Article{Behler2011,
  author    = {Behler, Jörg},
  journal   = {J. Chem. Phys.},
  title     = {Atom-centered symmetry functions for constructing high-dimensional neural network potentials},
  year      = {2011},
  issn      = {1089-7690},
  month     = feb,
  number    = {7},
  pages     = {074106},
  volume    = {134},
  doi       = {10.1063/1.3553717},
  fjournal  = {The Journal of Chemical Physics},
  publisher = {AIP Publishing},
}

@Article{Abrahamsson2009,
  author    = {Abrahamsson, Erik and Andersson, Stefan and Marković, Nikola and Nyman, Gunnar},
  journal   = {J. Phys. Chem. A},
  title     = {Dynamics of the O + CN Reaction and N + CO Scattering on Two Coupled Surfaces},
  year      = {2009},
  issn      = {1520-5215},
  month     = oct,
  number    = {52},
  pages     = {14824--14830},
  volume    = {113},
  comment   = {ISC modelling},
  doi       = {10.1021/jp904954k},
  fjournal  = {The Journal of Physical Chemistry A},
  publisher = {American Chemical Society (ACS)},
}

@Article{Fu2011,
  author    = {Fu, Bina and Shepler, Benjamin C. and Bowman, Joel M.},
  journal   = {JACS},
  title     = {Three-State Trajectory Surface Hopping Studies of the Photodissociation Dynamics of Formaldehyde on ab Initio Potential Energy Surfaces},
  year      = {2011},
  issn      = {1520-5126},
  month     = apr,
  number    = {20},
  pages     = {7957--7968},
  volume    = {133},
  comment   = {ISC modelling},
  doi       = {10.1021/ja201559r},
  fjournal  = {Journal of the American Chemical Society},
  publisher = {American Chemical Society (ACS)},
}

@Article{Han2011,
  author    = {Han, Boran and Zheng, Yujun},
  journal   = {J. Comput. Chem.},
  title     = {Nonadiabatic quantum dynamics in O(3P)+H2→OH+H: A revisited study},
  year      = {2011},
  issn      = {1096-987X},
  month     = sep,
  number    = {16},
  pages     = {3520--3525},
  volume    = {32},
  doi       = {10.1002/jcc.21940},
  fjournal  = {Journal of Computational Chemistry},
  publisher = {Wiley},
}

@Article{Li2009,
  author    = {Li, Bin and Han, Ke-Li},
  journal   = {J. Phys. Chem. A},
  title     = {Mixed Quantum-Classical Study of Nonadiabatic Dynamics in the O(3P2,1,0,1D2) + H2 Reaction},
  year      = {2009},
  issn      = {1520-5215},
  month     = sep,
  number    = {38},
  pages     = {10189--10195},
  volume    = {113},
  comment   = {ISC modelling},
  doi       = {10.1021/jp904727d},
  fjournal  = {The Journal of Physical Chemistry A},
  publisher = {American Chemical Society (ACS)},
}

@Article{Hu2007,
  author    = {Hu, Wenfang and Lendvay, György and Maiti, Biswajit and Schatz, George C.},
  journal   = {J. Phys. Chem. A},
  title     = {Trajectory Surface Hopping Study of the O(3P) + Ethylene Reaction Dynamics},
  year      = {2007},
  issn      = {1520-5215},
  month     = dec,
  number    = {10},
  pages     = {2093--2103},
  volume    = {112},
  comment   = {ISC modelling},
  doi       = {10.1021/jp076716z},
  fjournal  = {The Journal of Physical Chemistry A},
  publisher = {American Chemical Society (ACS)},
}

@Article{Curchod2016,
  author    = {Curchod, Basile F. E. and Rauer, Clemens and Marquetand, Philipp and González, Leticia and Martínez, Todd J.},
  journal   = {J. Chem. Phys.},
  title     = {Communication: GAIMS—Generalized Ab Initio Multiple Spawning for both internal conversion and intersystem crossing processes},
  year      = {2016},
  issn      = {1089-7690},
  month     = mar,
  number    = {10},
  pages     = {101102},
  volume    = {144},
  comment   = {ISC modelling},
  doi       = {10.1063/1.4943571},
  fjournal  = {The Journal of Chemical Physics},
  publisher = {AIP Publishing},
}

@Article{Fedorov2016,
  author    = {Fedorov, Dmitry A. and Pruitt, Spencer R. and Keipert, Kristopher and Gordon, Mark S. and Varganov, Sergey A.},
  journal   = {J. Phys. Chem. A},
  title     = {Ab Initio Multiple Spawning Method for Intersystem Crossing Dynamics: Spin-Forbidden Transitions between 3B1 and 1A1 States of GeH2},
  year      = {2016},
  issn      = {1520-5215},
  month     = apr,
  number    = {18},
  pages     = {2911--2919},
  volume    = {120},
  comment   = {ISC modelling},
  doi       = {10.1021/acs.jpca.6b01406},
  fjournal  = {The Journal of Physical Chemistry A},
  publisher = {American Chemical Society (ACS)},
}

@Article{Zaari2015,
  author    = {Zaari, Ryan R. and Varganov, Sergey A.},
  journal   = {J. Phys. Chem. A},
  title     = {Nonadiabatic Transition State Theory and Trajectory Surface Hopping Dynamics: Intersystem Crossing Between 3B1 and 1A1 States of SiH2},
  year      = {2015},
  issn      = {1520-5215},
  month     = feb,
  number    = {8},
  pages     = {1332--1338},
  volume    = {119},
  doi       = {10.1021/jp509515e},
  fjournal  = {The Journal of Physical Chemistry A},
  publisher = {American Chemical Society (ACS)},
}

@Article{Takayanagi2002,
  author    = {Takayanagi, Toshiyuki},
  journal   = {J. Phys. Chem. A},
  title     = {Quantum Scattering Calculations of the O($^1$D) + N$_2$(X$^1\Sigma_g^+$) $\rightarrow$ O($^3$P) + N$_2$(X$_1\Sigma_g^+$) Spin-Forbidden Electronic Quenching Collision},
  year      = {2002},
  issn      = {1520-5215},
  month     = apr,
  number    = {19},
  pages     = {4914--4921},
  volume    = {106},
  comment   = {ISC modelling},
  doi       = {10.1021/jp0200425},
  fjournal  = {The Journal of Physical Chemistry A},
  publisher = {American Chemical Society (ACS)},
}

@Article{Dreuw2026,
  author    = {Dreuw, Andreas},
  journal   = {Adv. Sci.},
  title     = {Why Computational Photochemistry Is Challenging and Will Probably Remain So: A Quantum Chemist’s Perspective},
  year      = {2026},
  issn      = {2198-3844},
  month     = mar,
  number    = {21},
  pages     = {e21012},
  volume    = {13},
  doi       = {10.1002/advs.202521012},
  fjournal  = {Advanced Science},
  publisher = {Wiley},
}

@Article{Dolmans2003,
  author    = {Dolmans, Dennis E.J.G.J. and Fukumura, Dai and Jain, Rakesh K.},
  journal   = {Nat. Rev. Cancer},
  title     = {Photodynamic therapy for cancer},
  year      = {2003},
  issn      = {1474-1768},
  month     = may,
  number    = {5},
  pages     = {380--387},
  volume    = {3},
  comment   = {PDT},
  doi       = {10.1038/nrc1071},
  fjournal  = {Nature Reviews Cancer},
  publisher = {Springer Science and Business Media LLC},
}

@Article{Agostinis2011,
  author    = {Agostinis, Patrizia and Berg, Kristian and Cengel, Keith A. and Foster, Thomas H. and Girotti, Albert W. and Gollnick, Sandra O. and Hahn, Stephen M. and Hamblin, Michael R. and Juzeniene, Asta and Kessel, David and Korbelik, Mladen and Moan, Johan and Mroz, Pawel and Nowis, Dominika and Piette, Jacques and Wilson, Brian C. and Golab, Jakub},
  journal   = {CA. Cancer J. Clin.},
  title     = {Photodynamic therapy of cancer: An update},
  year      = {2011},
  issn      = {0007-9235},
  month     = may,
  number    = {4},
  pages     = {250--281},
  volume    = {61},
  comment   = {PDT},
  doi       = {10.3322/caac.20114},
  fjournal  = {CA: A Cancer Journal for Clinicians},
  publisher = {Wiley},
}

@Article{Wang2022,
  author    = {Wang, Zhihang and Hölzel, Helen and Moth-Poulsen, Kasper},
  journal   = {Chem. Soc. Rev.},
  title     = {Status and challenges for molecular solar thermal energy storage system based devices},
  year      = {2022},
  issn      = {1460-4744},
  number    = {17},
  pages     = {7313--7326},
  volume    = {51},
  doi       = {10.1039/d1cs00890k},
  fjournal  = {Chemical Society Reviews},
  publisher = {Royal Society of Chemistry (RSC)},
}

@Article{Lowrie2023,
  author    = {Lowrie, William and Westbrook, Robert J. E. and Guo, Junjun and Gonev, Hristo Ivov and Marin-Beloqui, Jose and Clarke, Tracey M.},
  journal   = {J. Chem. Phys.},
  title     = {Organic photovoltaics: The current challenges},
  year      = {2023},
  issn      = {1089-7690},
  month     = mar,
  number    = {11},
  pages     = {110901},
  volume    = {158},
  doi       = {10.1063/5.0139457},
  fjournal  = {The Journal of Chemical Physics},
  publisher = {AIP Publishing},
}

@Article{Plasser2025,
  author    = {Plasser, Felix and Lischka, Hans and Shepard, Ron and Szalay, Péter G. and Pitzer, Russell M. and Alves, Rodolpho L. R. and Aquino, Adelia J. A. and Autschbach, Jochen and Barbatti, Mario and Carvalho, Jhonatas R. and Chagas, Julio C. V. and González, Leticia and Hansen, Andreas and Jayee, Bhumika and Kertesz, Miklos and Machado, Francisco B. C. and Matsika, Spiridoula and do Monte, Silmar A. and Mukherjee, Saikat and Nachtigallová, Dana and Nieman, Reed and Oliveira, Vytor P. and Oppel, Markus and Parish, Carol A. and Pittner, Jiri and F. dos Santos, Luan G. and Scrinzi, Armin and Sit, Mahesh K. and Spada, Rene F. K. and Thodika, Mushir and Valente, Daniel C. A. and Vázquez-Mayagoitia, Álvaro and Ventura, Elizete and Westermayr, Julia and Zaichenko, Aleksandr and Zhang, Zhiyong},
  journal   = {J. Phys. Chem. A},
  title     = {COLUMBUS--An Efficient and General Program Package for Ground and Excited State Computations Including Spin--Orbit Couplings and Dynamics},
  year      = {2025},
  issn      = {1520-5215},
  month     = jul,
  number    = {28},
  pages     = {6482--6517},
  volume    = {129},
  doi       = {10.1021/acs.jpca.5c02047},
  fjournal  = {The Journal of Physical Chemistry A},
  publisher = {American Chemical Society (ACS)},
}

@Misc{Hu2025,
  author    = {Hu, Jinming and Nawaz, Hassan and Hou, Yi-Fan and Rui, Yuting and Chi, Lijie and Chen, Yuxinxin and Ullah, Arif and Dral, Pavlo O.},
  title     = {Aitomia: Your Intelligent Assistant for AI-Driven Atomistic and Quantum Chemical Simulations},
  journal   = {arXiv:2505.08195v4 [physics.comp-ph] },
  year      = {2025},
  copyright = {Creative Commons Attribution 4.0 International},
  doi       = {10.48550/ARXIV.2505.08195},
  publisher = {arXiv},
}

@Article{Westermayr2020a,
  author    = {Julia Westermayr and Felix A Faber and Anders S Christensen and O Anatole von Lilienfeld and Philipp Marquetand},
  journal   = {Mach. Learn.: Sci. Technol.},
  title     = {Neural Networks and Kernel Ridge Regression for Excited States Dynamics of $CH_2NH_2^+$: From Single-State to Multi-State Representations and Multi-Property Machine Learning Models},
  year      = {2020},
  month     = {may},
  number    = {2},
  pages     = {025009},
  volume    = {1},
  comment   = {scaled NACs, reported difficulties in the ML-prediction of nonadiabatic couplings},
  doi       = {10.1088/2632-2153/ab88d0},
  fjournal  = {Machine Learning: Science and Technology},
  publisher = {{IOP} Publishing},
}

@Article{Mueller2025,
  author    = {M\"{u}ller, Carolin and Sr\v{s}e\v{n}, \v{S}t\v{e}p\'{a}n and Bachmair, Brigitta and Crespo-Otero, Rachel and Li, Jingbai and Mausenberger, Sascha and Pinheiro, Max and Worth, Graham and Lopez, Steven A. and Westermayr, Julia},
  journal   = {Chem. Sci.},
  title     = {Machine learning for nonadiabatic molecular dynamics: best practices and recent progress},
  year      = {2025},
  issn      = {2041-6539},
  month     = sep,
  pages     = {17542--17567},
  volume    = {16},
  doi       = {10.1039/d5sc05579b},
  fjournal  = {Chemical Science},
  issue     = {38},
  publisher = {Royal Society of Chemistry (RSC)},
}

@Article{NewtonX2026,
  author   = {Mario Barbatti and Rafael S. Mattos and Baptiste Demoulin and Matheus de O. Bispo and Mattia Bondanza and Marcus Brady and Rachel Crespo-Otero and Ely G. F. de Miranda and Pavlo O. Dral and Giovanni Granucci and Anna Hehn and Federico J. Hernández and Gabriele Iuzzolino and Ritama Kar and Fábris Kossoski and Hans Lischka and Benedetta Mennucci and Saikat Mukherjee and Anik Mukhopadhyay and Fulvio Perrella and Maurizio Persico and Max Pinheiro Jr and Jiri Pittner and Felix Plasser and Nadia Rega and Eduarda Sangiogo-Gil and Tejas Thorat and Josene M. Toldo and Anderson A. Tomaz and Márcio T. do N. Varella and Luis Vasquez},
  journal  = {ChemRxiv},
  title    = {The Newton-X Platform for Mixed Quantum--Classical Dynamics},
  year     = {2026},
  number   = {0419},
  volume   = {2026},
  doi      = {10.26434/chemrxiv.15002063/v2},
  eprint   = {https://chemrxiv.org/doi/pdf/10.26434/chemrxiv.15002063/v2},
  url      = {https://chemrxiv.org/doi/abs/10.26434/chemrxiv.15002063/v2},
}

@Article{NewtonX2014,
  author    = {Barbatti, Mario and Ruckenbauer, Matthias and Plasser, Felix and Pittner, Jiri and Granucci, Giovanni and Persico, Maurizio and Lischka, Hans},
  journal   = {WIREs Comput. Mol. Sci.},
  title     = {Newton‐X: a surface‐hopping program for nonadiabatic molecular dynamics},
  year      = {2014},
  issn      = {1759-0884},
  month     = jun,
  number    = {1},
  pages     = {26--33},
  volume    = {4},
  comment   = {old Newton-X},
  doi       = {10.1002/wcms.1158},
  fjournal  = {WIREs Computational Molecular Science},
  publisher = {Wiley},
}

@Article{Barbatti2010b,
  author    = {Barbatti, Mario and Aquino, Adelia J. A. and Lischka, Hans},
  journal   = {Phys. Chem. Chem. Phys.},
  title     = {The UV absorption of nucleobases: semi-classical ab initio spectra simulations},
  year      = {2010},
  issn      = {1463-9084},
  number    = {19},
  pages     = {4959},
  volume    = {12},
  doi       = {10.1039/b924956g},
  fjournal  = {Physical Chemistry Chemical Physics},
  publisher = {Royal Society of Chemistry (RSC)},
}

@Article{WasifBaig2021,
  author    = {Wasif Baig, Mirza and Pederzoli, Marek and Kývala, Mojmír and Cwiklik, Lukasz and Pittner, Jiří},
  journal   = {J. Phys. Chem. B},
  title     = {Theoretical Investigation of the Effect of Alkylation and Bromination on Intersystem Crossing in BODIPY-Based Photosensitizers},
  year      = {2021},
  issn      = {1520-5207},
  month     = oct,
  number    = {42},
  pages     = {11617--11627},
  volume    = {125},
  doi       = {10.1021/acs.jpcb.1c05236},
  fjournal  = {The Journal of Physical Chemistry B},
  publisher = {American Chemical Society (ACS)},
}

@InProceedings{Batatia2022,
  author    = {Batatia, Ilyes and Kovács, Dávid Péter and Simm, Gregor N. C. and Ortner, Christoph and Csányi, Gábor},
  title     = {MACE: Higher Order Equivariant Message Passing Neural Networks for Fast and Accurate Force Fields},
  booktitle = {Advances in Neural Information Processing Systems},
  year      = {2022},
  pages     = {11423--11436},
  volume    = {35},
  comment   = {MACE},
  publisher = {Curran Associates, Inc.},
  editor    = {S. Koyejo and S. Mohamed and A. Agarwal and D. Belgave and K. Cho and A. Oh},
}

@InProceedings{Schutt2021,
  author    = {Kristof T. Sch{\"u}tt and Oliver T. Unke and Michael Gastegger},
  booktitle = {Proceedings of the 38th International Conference on Machine Learning},
  title     = {Equivariant message passing for the prediction of tensorial properties and molecular spectra},
  year      = {2021},
  pages     = {9377--9388},
  volume    = {139},
  editor    = {Meila, Marina and Zhang, Tong},
  month     = jul,
  url       = {https://arxiv.org/abs/2102.03150},
  series    = {Proceedings of Machine Learning Research},
  publisher = {PMLR},
}

@Article{Bakulin2012,
  author    = {Bakulin, Artem A. and Rao, Akshay and Pavelyev, Vlad G. and van Loosdrecht, Paul H. M. and Pshenichnikov, Maxim S. and Niedzialek, Dorota and Cornil, Jérôme and Beljonne, David and Friend, Richard H.},
  journal   = {Sci.},
  title     = {The Role of Driving Energy and Delocalized States for Charge Separation in Organic Semiconductors},
  year      = {2012},
  issn      = {1095-9203},
  month     = mar,
  number    = {6074},
  pages     = {1340--1344},
  volume    = {335},
  doi       = {10.1126/science.1217745},
  fjournal  = {Science},
  publisher = {American Association for the Advancement of Science (AAAS)},
}

@Article{Schuett2018a,
  author    = {Schütt, K. T. and Sauceda, H. E. and Kindermans, P.-J. and Tkatchenko, A. and Müller, K.-R.},
  journal   = {J. Chem. Phys.},
  title     = {SchNet – A deep learning architecture for molecules and materials},
  year      = {2018},
  issn      = {1089-7690},
  month     = mar,
  number    = {24},
  pages     = {241722},
  volume    = {148},
  comment   = {SchNet},
  doi       = {10.1063/1.5019779},
  fjournal  = {The Journal of Chemical Physics},
  publisher = {AIP Publishing},
}

@Book{Efron1994,
  author    = {Efron, Bradley and Tibshirani, R.J.},
  publisher = {Chapman and Hall/CRC},
  title     = {An Introduction to the Bootstrap},
  year      = {1994},
  isbn      = {9780429246593},
  month     = May,
  doi       = {10.1201/9780429246593},
}

@Book{Mooney1993,
  author    = {Christopher Z. Mooney and Robert D. Duval},
  publisher = {SAGE},
  title     = {Bootstrapping: A Nonparametric Approach to Statistical Inference},
  year      = {1993},
  address   = {Newbury Park, CA, USA},
  comment   = {Bootstrapping},
}

@Article{Efron1979,
  author    = {Efron, B.},
  journal   = {The Annals of Statistics},
  title     = {Bootstrap Methods: Another Look at the Jackknife},
  year      = {1979},
  issn      = {0090-5364},
  month     = Jan,
  number    = {1},
  volume    = {7},
  comment   = {Bootstrap},
  doi       = {10.1214/aos/1176344552},
  publisher = {Institute of Mathematical Statistics},
}

@Article{Lorquet1988,
  author    = {Lorquet, J. C. and Leyh-Nihant, B.},
  journal   = {The Journal of Physical Chemistry},
  title     = {Nonadiabatic unimolecular reactions. 1. A statistical formulation for the rate constants},
  year      = {1988},
  issn      = {1541-5740},
  month     = Aug,
  number    = {16},
  pages     = {4778--4783},
  volume    = {92},
  comment   = {nonadiabatic transition state theory},
  doi       = {10.1021/j100327a043},
  publisher = {American Chemical Society (ACS)},
}

@InBook{SOC1977,
author="S. R. Langhoff and C. W. Kern",
pages={381-437},
publisher="Plenum Press",
booktitle="Applications of Electronic Structure Theory",
editor="H. F. Schaefer~III",
year=1977,
address="New York"
}



\end{document}